\RequirePackage{fix-cm}
\documentclass{svjour3}                     
\smartqed  
\usepackage{graphicx}
\usepackage{amsmath}
\usepackage{amsfonts}
\usepackage{amssymb}
\usepackage{graphicx}
\usepackage{mathrsfs}
\usepackage{siunitx}
\usepackage{array}
\newcommand{\sign}{\text{sign}}
\newcommand{\ud}[1]{\, \mathrm{d}#1}
\newcommand{\deriv}[3][]{\frac{\ud^{#1} \hspace{-0.3mm} #2}{\ud{#3}^{#1}}}
\newcommand{\pderiv}[3][]{\frac{\d^{#1} \hspace{-0.1mm} #2}{\d{#3}^{#1}}}

\newcommand{\grad}[1][]{\nabla_{\! #1} }

\newcommand{\EXP}{\mathbb{E}}

\newcommand {\bs}[1]{\boldsymbol{#1}}

\newcommand{\vp}{\varphi}

\newcommand{\om}{\omega}

\newcommand{\R}{\mathbb{R}}
\renewcommand{\geq}{\geqslant}

\renewcommand{\leq}{\leqslant}
\renewcommand{\d}{\partial}

\newcommand{\ub}{\boldsymbol{u}}
\newcommand{\xb}{\boldsymbol{x}}

\begin{document}

\title{Modeling tropotaxis in ant colonies: recruitment and trail formation}


%


\titlerunning{Tropotaxis in ant colonies}        

\author{Jorge M. Ramirez         \and Sara M. Vallejo-Bernal \and Yurani Villa \and Sara Gaona \and Sarai Quintero
}


\institute{Universidad Nacional de Colombia, Sede Medell\'in \\
              Tel.: +57-4-4309000 Ext 46313 \\
              \email{jmramirezo@unal.edu.co}           
}

\date{Received: date / Accepted: date}

\maketitle

\begin{abstract}
We propose an active walker model for the motion of individual ants communicating via chemical signals. It is assumed that communication takes the form of a time-dependent pheromone field that feedbacks into the motion ants through tropotaxis: individuals can sense the gradient of the pheromone concentration field and adjust their orientation accordingly. The individual model takes the form of a Langevin system of equations in polar coordinates driven by two-dimensional Gaussian fluctuations and with orientation changes in response to two pheromone fields: one emanating from the nest, and other actively produced by ants in their nest-bound journey after finding a food source. We explicitly track the evolution of both fields in three dimensions. The proposed tropotaxis model relating the pheromone field to the orientation changes is similar to Weber's law, but depends explicitly only on the gradient of the pheromone concentration. We identify ranges of values for the model parameters that yield the emergence of two key foraging patterns: successful recruitment to newly found sources, and colony-wide trail networks.   
\end{abstract}

\section{Introduction}\label{Sec_Introduction}

As thoroughly detailed in \cite{holldobler1990ants}, the approximately 8,800 known ant species use a dazzling variety of procedures for foraging and retrieving food collaboratively. Similarly diverse are their communication and navigation strategies, which include the use of chemical, tactile, visual and acoustic signals with a wide scope of functional responses.   Moreover, a single colony of ants may practice different foraging, orientation and communication techniques simultaneously. Within such diversity, myrmecologists have long identified chemical signals, and pheromones in particular, as the central mechanism in the organization of ant societies. Chemical communication is as diverse as it is pervasive: researchers have identified tenths of kinds of signals, corresponding to several types of chemicals produced in as much as ten distinct organs, associated with a plethora of adaptations.

There is thus no single appropriate answer to the question of how ants forage, recruit or communicate. A different question, and one mathematical analyses may help answer, is under what conditions some specific set of individual behavioral rules may  reproduce well-known colony self-organization patterns and dynamics. Namely, by appropriately up-scaling individual based models to the colony scale, we can identify necessary conditions on the assumed models for the emergence of observed patterns, and thus propose testable hypotheses on the corresponding behavioral traits at the individual level.

We set out on this task by considering an idealized colony of ants bound to a single nest, where foraging, recruitment and orientation are mediated solely by chemical communication. We are particularly interested in how single ants, by adequately marking the ground with pheromones, can potentially recruit large numbers of nest-mates to a food source. These pheromones, along with a specific nest odor used for homeward orientation, comprise a time-evolving and self-generated odor landscape which ants use to dynamically orient their motion \cite{holldobler1990ants,wyatt2003pheromones,detrain2008collective}. 

Among all possible mechanisms for animal orientation towards or away from a chemical stimulus, `tropotaxis' has been identified as the most basic one used by ants and termites in general, and in particular, fundamental for trail following \cite{hangartner1967spezifitat,leuthold1975orientation}. As defined in \cite{wyatt2003pheromones}, tropotaxis is ``the ability of performing directed turns based on simultaneous comparison of stimulus intensity on two points of the body". Ants, in particular, as they travel through the odor landscape, compare stimulus inputs from left and right antennae, computing the concentration difference and changing direction towards the direction of higher concentration \cite{hangartner1967spezifitat,holldobler1990ants,calenbuhr1992model}.

Whether a chemical signal reaches an ant and communication takes place, depends on the dynamics of three-dimensional molecular diffusion of pheromones through air. Specifically, as noted in \cite{holldobler1990ants} and \cite{wyatt2003pheromones}, ants following a trail respond to the airborne molecules of pheromone diffusing from the marks, rather than with a contact response to the trail itself. As pheromones diffuse through air from the ground-level sources, an `active space' of communication is created, this is the zone within which the concentration of a pheromone is at or above a concentration threshold, and where tropotaxis can occur \cite{bossert1963analysis}.

The exact nature of tropotaxis, namely of how individual ants convert pheromone concentration differences into orientation changes, has been a source of lively research both by experimentalist and modelers alike (see \cite{camazine2003self} and \cite{amorim2018ant} for a review). Although there is no reason to expect a single mechanism to cover all diversity in the ant family, experimental and analytical studies pioneered by \cite{perna2012individual} point to `Weber’s law' as a plausible explanation. Namely, it is claimed that the turning rate is determined by the difference in pheromone on both sides of the ant, divided by the sum of pheromone on each side. Written in its simplest differential form, this means that the rate at which ants change their orientation $ \varphi $, in response to a pheromone field of concentration $ F $, is $ \ud \vp $ proportional to $\grad F/F$, or equivalently $ \ud \vp \propto \grad(\log F) $. Recently in \cite{amorim2018ant}, the authors  perform a mathematical analysis on a generalization of Weber's law, and give conditions on its parameters under which the formation of straight trails emerges as a stable pattern. 

In this work we consider and individual based model with a tropotaxis formulation similar to (but simpler than) Weber's law,  and study the emergence of a number of observed colony-wide patterns, including trail formation. Specifically, we suppose that $ \ud \vp \propto \log (|\grad F|) $ and establish conditions under which a colony of ants is able to: (1) forage within an area of given size around a central nest, (2) recruit ants to newly-discovered food sources and (3) form a complex pattern of trails emanating from the nest. Our framework also explicitly includes randomness in the equation of ant motion, and the detailed mechanics of pheromone diffusion. In our case this is done by a combination of mathematical analysis and simulations.

We build upon a rich modeling tradition of ant behavior dating as far back as the work of \cite{bossert1963analysis}, and culminating in the recent work by \cite{amorim2018ant}. In particular, we note the work of \cite{ryan2016model} from which we draw heavily. As in the latter study, we couple an \textit{Active Walker Model} for the motion of individual ants, with a model for the spatio-temporal evolution of the pheromone fields. In \cite{lam1995active,helbing1997active,schweitzer1997active} the authors use active walker models to study trail formation but disregard molecular diffusion of the pheromone field. In \cite{perna2012individual} the authors make a key finding: ants generally respond to the pheromone stimuli by modifying their orientation, not their speed. This feature is not included in \cite{ryan2016model} but is explicitly incorporated in our model. \cite{boissard2013trail} uses a jump process for the orientation of individual ants but does not consider tropotaxis as the feedback mechanism. In \cite{taktikos2012collective} a model similar to ours is proposed but the authors do not address any questions on recruitment or trail formation.

Our approach shares a similar intention with many of the studies cited above. That is, we propose a bio-physically sensible individual based model and study its properties, specifically looking for regions of its parameter space that guarantee the emergence of known colony-wide behavior and patterns. This does not constitute fitting parameter values in order to obtain expected results. Rather, by identifying necessary conditions on the parameters of the equations for ant motion, we may draw conclusions or posit hypotheses on the individual behavioral traits that underlie the emergence of large scale patterns. 

The first part of this paper concerns the formulation of the model. In Section \ref{Sec_Model} we describe the behavioral assumptions of the model followed by their corresponding mathematical description in Section \ref{Sec_MathModel}. There, we lay out the Langevin equations of motion for individual ants as well as the equations governing the production and evolution of the pheromone fields. All equations are first constructed in dimensional form, and following baseline estimates on a handful of variables, casted into a non-dimensional model in Section \ref{Sect_adim}. In Section \ref{Sec_Tropotaxis} we describe our mathematical model for tropotaxis. Namely we formulate equations of the magnitude and direction of the rate at which ants change their orientation as a function of a pheromone field. There, we also propose a parametric pheromone-dependent noise coefficient that can produce qualitatively appropriate exploration trails.

Once the model is complete, we move on to study conditions under which recruitment and trail patterns emerge. Section \ref{Sec_recruitment} introduces the conditions for \textit{successful recruitment}: the ability of an ant that has just found a food source, to leave pheromone markings along its nest-bound path such that any ant located at the nest can follow the pheromone signal to the food source. We find that successful recruitment necessitates very special values for the pheromone marking rate, as well as very high sensitivity to pheromone gradients. Section \ref{Sec_trails} describes two results regarding the formation and structure of colony-wide trail networks. First we find that the connectivity of the resulting trail system can be controlled by a parameter that directs ant motion to previously marked trails on their nest bound path. Secondly we note that our model reproduces the observed switch from multiple trails of few ants to few trails of many ants \cite{pasteels1987self,deneubourg1989blind,detrain2008collective}, and thus can be explained by tropotaxis alone.

\section{A tropotaxis model for ant foraging and recruitment}\label{Sec_Model}

We consider a fixed number of ants foraging in two-dimensional space around the vicinity of a fixed nest. Ants follow a set of well-established behavioral rules which we now summarize, (see for example \cite{holldobler1990ants,bossert1963analysis,beckers1992trails}). The mathematical formulation of the model is derived in  subsequent sections. 

We consider only two possible modes for any ant: \textit{foraging} and \textit{recruitment}. During the foraging mode, an ant will search for food adjusting its orientation guided by the pheromone field resulting from ground markings. Whenever the ant reaches a food source, she enters into recruitment mode. She then returns to the nest leaving a trail of pheromone markings with the objective that others can follow it to the food source she just found. When the recruiting ant reaches the nest, she becomes a forager again and her routine starts all over. We call this the \textit{foraging-recruitment cycle}.

The movement of each individual ant is determined by three time-dependent state variables: \textit{position}, \textit{speed} and \textit{orientation}. Most of our modeling efforts are directed at describing the dynamics of the orientation, which is defined as the angle spanned between the head-to-gaster axis of an ant, and an arbitrary but fixed reference direction. The velocity vector of each ant points along her orientation and has magnitude equal to her speed. The ant's position, which is measured with respect to the nest, is simply obtained by integrating the velocity vector over time. 

In our model, the movement of each ant is driven by two underlying processes: a \textit{field of pheromone concentration}, and \textit{random fluctuations}.  The pheromone field forms a time-evolving odour landscape that the ants use to dynamically orient their motion \cite{wyatt2003pheromones}. A crucial simplification is that ants respond to the pheromone field by modifying their orientation, not their speed. This is based on the observations made by \cite{perna2012individual} and is the main motivation for us to propose a model based only on tropotaxis: by sensing the strength and gradient of the pheromone field at its location, the ant takes small turns thus adjusting its orientation in order to move in the direction of largest concentration gradient. See also \cite{schienbein1993langevin,taktikos2012collective,amorim2018ant}.

Molecular diffusion of pheromones in air is the underlying physical mechanism on which communication by tropotaxis depends: from a point source, pheromones diffuse outwards creating a signal in the form of a hemispherical `active space' where concentration exceeds a threshold and sensing can take place \cite{holldobler1990ants,wyatt2003pheromones}. The mathematical analysis of the dynamics of pheromone diffusion and its relation to recruiting in ants was pioneered by \cite{bossert1963analysis}. The comprehensive work by  \cite{wyatt2003pheromones} details the molecular structure of a plethora of pheromones found in ants, and how they relate to diffusion rates in air. 

Two distinct sources of pheromone are considered: 
\begin{enumerate}
\item the nest, which acts as a beacon producing a constant pheromone field from its point source, and
\item the pheromone signal produced by recruiting ants as they mark the ground along the path on their way to the nest. 
\end{enumerate}

Some remarks are in order. We are supposing that pheromone fields are the sole means of navigation and communication among ants. Namely we assume ants are blind, use no spatial or magnetic cues, cannot measure distances and do not interact physically with others. The proposed existence of a `nest pheromone' for navigation falls in line of observations of some ant species where workers mark the substrate in the vicinity of their nest entrances and use the odor to orient homeward, or even the simpler practice of moving up carbon dioxide gradients, and hence in the direction of the largest nearby clusters of ants \cite{holldobler1990ants,deneubourg1989blind}. In general however, assuming directed turns towards the single maximum of a static pheromone field can also be understood as a model for ants adjusting their orientation in the direction of a known fixed position, whatever the sensory nature of this knowledge may be. 

An important underlying assumption of our model is that no directional information is included in the pheromone markings, namely there is no \textit{trail polarization}. As noted in \cite{holldobler1990ants}, for the trail-following species of the genus \textit{Lasius} there is conclusive evidence that trail polarization does not occur. It has however been observed in ants of the genera \textit{Myrmica} and \textit{Monomorium} \cite{jackson2004trail}. Moreover, trail polarization has been successfully modeled to predict the evolution and formation of trail patterns in \cite{boissard2013trail}.

Noise enters our model as Gaussian random increments in the equations for the dynamics of the ant's speed and orientation. It is not only used to capture the apparently erratic and inherently uncertain motion of individual ants, but also as means to include the uncertainty associated with the fact that our model ignores several processes whose compounded effect we assume is small. These processes are specific-dependent and might include:  micro-topography, ant-ant collisions, visual cues, other pheromone signals, etcetera. See for example \cite{sudd2013behavioural,schweitzer2007brownian}. For stochastic formulations in terms of distributions other than Gaussian, see \cite{boissard2013trail} and \cite{cheung2009mathematical}.

We introduce the notion of \textit{successful recruitment} to denote the ability of a recruiter ant to leave marks that will guide a forager ant from the nest to a food source. To be precise, \textit{successful recruitment} requires that the pheromone markings left by the recruiter on its path satisfy the following conditions: 
\begin{enumerate}
\item  there is a path connecting the nest and the food source along which the pheromone field is increasing,
\item the pheromone concentration is above the perception threshold at all points along such trail,  
\item conditions 1 and 2 hold for a long enough amount of time, so ants have the opportunity to reach the food source from the nest or other nearby locations.
\end{enumerate}

The existence of a strategy of successful recruitment is necessary for the emergence of colony-scale trail patterns of food retrieval. The geometry and definiteness of such patterns will depend on \textit{trail preference}. Namely, recruiter ants might not follow the shortest path in return the nest after finding a food source, but will rather follow existing trains of moving ants \cite{wyatt2003pheromones}. 

In subsequent sections we formulate the mathematical model for the processes outlined above. After picking characteristic scales, the model is cast in non-dimensional form, and some of its parameters are tuned in order to arrive at conditions of \textit{successful recruitment}. Finally, we perform numerical explorations on the role that trail preference play in the formation of realistic trail patterns.

\section{Mathematical formulation}\label{Sec_MathModel}

We propose an \textit{Active Walker Model} in which ants move over a flat two-dimensional surface interacting and modifying their environment through the production of pheromones. The evolution of the dynamical variables of each ant is specified via a system of ordinary and stochastic differential equations, coupled with partial differential equations modeling the evolution of the pheromone field ants produce and respond to. For a review of active walker models and their relevance in the study of trail formation, we refer the reader to \cite{helbing1997active}.

Below, for each relevant variable, we specify its units in the CGS system and write down the dimensional equations governing its behavior, as well as all parameters involved. In the next section, we will identify scales of interest and derive the non-dimensional counterparts.  

Note that no attempt is made here at associating our analysis with those of a particular ant species or family. Consequently the `data' used below is regarded mostly as baseline values that set the problem's spatio-temporal scales. Much of the subsequent work is then aimed at finding parameter values that would reproduce commonly observed behavior in ant colonies. All of our results are qualitative in nature, as is to be expected from a conceptual model.

\subsection{Equations of motion}

\begin{figure}[h]\centering
	\includegraphics[scale=1]{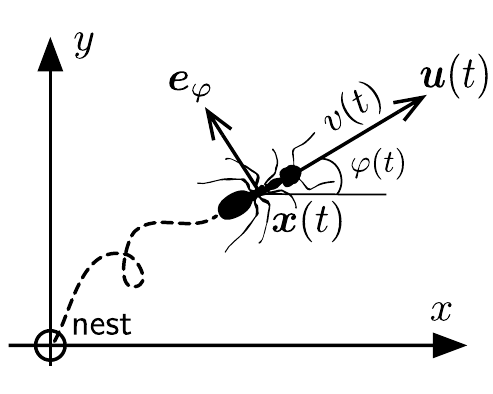}
	\caption{Schematic representation of the coordinate system and dynamic variables associated with each ants movement \label{Fig_Polar}}
\end{figure}

The position of an ant at time $t \geq 0$ (\si{s}) is $\xb(t) = (x(t),y(t))\in \R^2$ (\si{cm}) measured with respect to arbitrarily oriented $xy$-axes with its origin $(0,0)$ fixed at the nest's location. Its velocity vector is $\ub(t)$ (\si{cm/s}). A basic model for the ants motion is the following Langevin system for and active walker:
\begin{equation}\label{Eq_Langevinu}
	\dot \xb = \ub, \quad \ud \ub = \left(-\lambda \ub + \mathcal{F}(\xb,\ub,t)\right) \ud t + \sqrt{2 D_v} \ud \boldsymbol{B}_{\ub}
\end{equation}
where $ \lambda $ (\si{s^{-1}}) is a constant friction rate, $ \mathcal{F} $ encodes the nonlinear feedback mechanism mediated by the pheromone fields, $D_v$ (\si{cm^2 / s^2/s}) is a diffusion coefficient, and $ \bs{B_u} $ (\si{s^{1/2}}) is a two-dimensional Weiner process. See  \cite{helbing1997active,ebeling2003active,romanczuk2012active,ryan2016model} for similar models. It is assumed here that the pheromone feedback operates only on the direction of movement and not on its speed. We thus decompose the velocity vector as
\begin{equation}\label{Def_u}
	\ub(t) = v(t) (\cos \vp(t), \sin \vp(t))
\end{equation}
where  $v(t)$ (\si{cm/s}) is the speed and  $\vp(t)$ (\si{rad}) is the ant's orientation defined as the counter-clock-wise angle its head-to-gaster axis forms with the $x$-axis. (See Figure \ref{Fig_Polar} and \cite{romanczuk2012active}).

By a slight modification of the analysis of \cite{vakeroudis2015windings}, one can derive from equations (\ref{Eq_Langevinu} - \ref{Def_u}) the following system of stochastic differential equations for the processes $ v(t),\vp(t) $:
\begin{align}
\ud v &= \left(\frac{D_v}{v}-\lambda v \right) \ud t + \sqrt{2 D_v} \ud B_v, \label{Eq_dv}\\
\ud \vp &=  \omega(\xb,\vp,t) \ud t  + \frac{\sqrt{2 D_v}}{v} \ud B_\vp \label{Eq_dphi} 
\end{align}
where $ B_v$ and $ B_\vp $ (\si{s^{1/2}}) are a pair of independent one-dimensional standard Wiener processes. The term $ \omega $ (\si{1/s}) in \eqref{Eq_dphi} is the manifestation of the pheromone feedback mechanism upon the orientation process and is denoted as such because it truly is an angular velocity: it gives the rate of change of the ant's orientation as a function of the local sensory input. It plays a starring role as the model for tropotaxis and  will be derived in Section \ref{Sec_Omega}.

It can be shown that the distribution of $ v(t) $ given by \eqref{Eq_dv}, converges to an invariant distribution with mean and variance given respectively by:
\begin{equation}
v^* = \sqrt{\frac{\pi D_v}{2 \lambda}}, \quad \sigma_v^* = \frac{(4-\pi) D_v}{2 \lambda}
\end{equation}
which could be used in the field to derive the appropriate values of $ \lambda, D_v $ form observations of a particular population. Moreover, this convergence is independent of the pheromone feedback mechanism, which implies that the process $ \{v(t): t\geq 0\}$ plays a secondary role in the emerging of large-scale foraging patterns, the focus of the present study. For the remaining of the paper, we thus replace the dynamics of $ v(t) $ by its limiting behavior and arrive at the final system of equations 
\begin{align}
	\dot \xb &= \ub \label{Eq_xdot}\\
	\ub(t) &= v^*(\cos \vp(t), \sin \vp(t)), \label{Eq_u}\\
	\ud \vp &=  \omega(\xb,\vp,t) \ud t  + \sqrt{2 D_\vp} \ud B_\vp, \label{Eq_dphi2}
\end{align}
where
\begin{equation}
	D_\vp = \frac{D_v}{(v^*)^2} \label{Def_Dphi}
\end{equation}
is the \textit{angular dispersion coefficient}  (\si{1/s}) resulting from making $ v = v^* $ and $ \sqrt{2 D_\vp} := \sqrt{2D_v}/v^* $ in \eqref{Eq_dphi}. More will be said about $ D_\vp $ in Section \ref{Sec_Dispersion}. 

The model is completed by coupling the dynamics of $ N $ walkers determined by the triplet $ \xb_j, \ub_j, \vp_j$ with $ j=1,\dots,N $, evolving according to $ N $ versions of equations (\ref{Eq_xdot}-\ref{Eq_dphi2}), each driven by and independent Brownian process $ B_{\vp_j} $. 

\subsection{The pheromone fields}
 
We suppose that ants move along the two-dimensional boundary of a three-dimensional half-space on which the pheromones undergo linear molecular diffusion. Thus, the pheromone fields of interest to the ant dynamics are produced and must be evaluated at the `ground level' $z=0$. Namely we are interested in the concentration $F(\xb,t) = \bs{F}((\xb,0),t)$ where $\bs{F}$ (\si{mol/cm^3}) is a scalar field that evolves according to following three-dimensional diffusion process in half space
\begin{gather}
\begin{aligned}
\pderiv{\bs{F}}{t} &= D_F \, \Delta_3 \bs{F} + \Phi(\xb,t) \delta_0(z), \quad \xb \in \R^2, \, z>0\\
\pderiv{\bs{F}}{z} &= 0, \quad \quad \xb \in \R^2, \, z=0,
\end{aligned}\label{Eq_F3D}
\end{gather}
where $\Delta_3$ denotes the Laplace operator with respect to $(\xb,z)=(x,y,z)$ and $D_F$ (\si{cm^2/seg}) is the pheromone diffusion coefficient. The Dirac delta function $\delta_0$ (\si{cm^{-1}}) indicates that the source of pheromone comes from ground level and at a rate $\Phi$ (\si{mol/cm^3/s}). Note that the second line in \eqref{Eq_F3D} declares the ground as a no-flux or reflective boundary for pheromone molecules. The solution to \eqref{Eq_F3D} is 
\begin{equation}\label{Eq_solF}
\bs{F}((\xb,z),t)=  \int_0^t \int_{\R^2} \int_0^\infty \Phi(\bs{\xi},s) \delta_0(\zeta) G_N((\bs{\xi},\zeta),(\xb,z),t-s) \ud \zeta \ud \bs{\xi} \ud s
\end{equation}
where $G_N$ is the Green's function associated with three-dimensional diffusion problem in half-space with Neumann boundary condition at $z=0$:
\begin{equation}\label{Eq_GN}
G_N((\bs{\xi},\zeta),(\xb,z),t) = \frac{1}{8(\pi D_F t)^{3/2}}  \exp \left[- \frac{\|\xb - \bs{\xi}\|^2}{4 D_F t} \right] \left(\exp\left[- \frac{(z - \zeta)^2}{4 D_F t} \right]+ \exp\left[- \frac{(z + \zeta)^2}{4 D_F t} \right]\right).
\end{equation}
(See for example \cite{polyanin2015handbook} Equation 3.2.1-2.) Performing the inner-most integral in \eqref{Eq_solF} and evaluating at $z=0$ yields the ground level pheromone as
\begin{equation}\label{Eq_Fsol}
	F(\xb,t) = \bs{F}((\xb,0),t)= \int_0^t \int_{\R^2} \Phi(\bs{\xi},s) G(\bs{\xi},\xb,t-s) \ud \bs{\xi} \ud s, \quad \xb \in \R^2, \, t>0,
\end{equation}
where the kernel $G$ (\si{cm^{-3}}) is given by
\begin{equation}\label{Def_G}
	G(\bs{\xi},\xb,t) = G_N((\bs{\xi},0),(\xb,0),t) = \frac{1}{4(\pi D_F t)^{3/2}}  \exp \left[- \frac{\|\xb - \bs{\xi}\|^2}{4 D_F t} \right]
\end{equation}
and represents the fraction of pheromone concentration that is diffused from $\bs{\xi}$ to $\xb$ (both locations in the plane) in time $t$.

Two pheromone fields are considered in our model: $F_N$ which emanates from the nest, and $F_A$ which is produced by recruiting ants. We will refer to these as the nest and ant pheromone fields, respectively. Both fields are subject to the dynamics of equation \eqref{Eq_Fsol} but with one important difference: the nest source $\Phi_N$ is constant in time and concentrated at the origin, while the ant source $\Phi_A$ is concentrated along multiple one-dimensional paths and its concentration varies along each path. 

The nest pheromone field $F_N$ is thus quite simple as we assume it comes from the stationary solution to \eqref{Eq_F3D} with a constant source of strength $f_N$ (\si{mol/seg}) located at the origin. Namely $\Phi_N = f_N \delta_{(0,0)}(\xb)$ in \eqref{Eq_Fsol}, which yields
\begin{equation}
F_N(\xb) = \lim_{t \to \infty} \int_0^t f_N \, G(\bs{0},\xb,t-s) \ud s = \frac{f_N}{2 \pi D_F \|\xb\|}.
\end{equation}

To arrive at an expression for $F_A$ we must take into account the contributions of all ants. Consider thus the $j$-th ant and suppose it starts in foraging mode at $t=0$. Let $T_{j,1}^F$ be the first time it reaches a food source. If it subsequently returns back to the nest at time $T_{j,1}^N$, it means that during $T_{j,1}^F < t < T_{j,1}^N$ the ant was in recruiting mode marking the ground with pheromones, which we assume does at a rate  
\begin{equation}\label{Def_gj}
g_j(t) = f_A \, e^{-\kappa t}, \quad T_{j,1}^F < t < T_{j,1}^N.
\end{equation}
Here, $\kappa$ (\si{1/s}) controls how fast the ant reservoir depletes, while $f_A$ (\si{mol/seg}) is the maximum  rate at which the ground is marked \cite{schweitzer1997active,helbing1997active,ryan2016model}. Two comments are in order about our choice for the marking rate function $g_j$: first of all, it is a decreasing non-negative function, which is necessary for the conditions of successful recruitment to hold.  Secondly, the form of $g_j$ is consistent with our assumption that ants cannot measure distances: they behave as emptying linear reservoirs of pheromone, leaving marks at a rate proportional to an assumed depleting amount of available pheromone in their glands. 

The history of the $j$-th ant's routine up to time $t$ can then be summarized by the sequence of times 
$$ 0 \leq  T_{j,1}^F < T_{j,1}^N < \cdots < T_{j,m}^F < T_{j,m}^N < \cdots < T_{j,M_j(t)}^{N} \leq t, $$
where $M_j(t)$ denotes the number of times the $j$-th ant has found distinct sources of food in between returns to the nest, and we prescribe $T_{j,M_j(t)}^{N}=t$ if the ant has not yet arrived to the nest by time $t$. Combining \eqref{Eq_Fsol}, \eqref{Def_gj} and summing over all ants, we arrive at the following expression for the ant pheromone field: 
\begin{equation}\label{Eq_FA}
	F_A(\xb,t) = \sum_{i=1}^N \sum_{m=1}^{M_i(t)} \int_{T_{i,m}^F}^{T_{i,m}^N} f_A e^{-\kappa(s-T^F_{i,m})} G(\xb_i(s),\xb,t-s) \ud s.
\end{equation}

\subsection{Typical scales and adimensionalization}\label{Sect_adim}

In our conceptual approach we do not intend to predict the behavior of any ant colony of a particular species under some definite conditions. That would require the estimation of values for $ v^*$, $ D_\vp$,  $D_F$, $f_N$, $ \kappa $ and  $f_A $ from field data which is most likely impossible. Our aim is to establish the existence of physically and physiologically meaningful ranges of values for said parameters, such that the model reproduces large-scale qualitative patterns observed in ant colony foraging. We thus selected from the literature baseline parameter values, shown in Table \ref{Table_dimVars}, and use them to arrive at a non-dimensional model.

\begin{table}[h!]\centering
\scalebox{0.9}{\begin{tabular}{|c|p{4cm}|c|p{5cm}|}\hline
	Symbol & Description &  Baseline value & Reference\\\hline
	$r^*$	& Maximum foraging radius & \SI{100}{cm} & \cite{bossert1963analysis,perna2012individual}\\\hline
	$v^*$	& Mean walking speed & \SI{1.5}{cm/s} & \cite{ryan2016model,amorim2015modeling}  \\\hline
	$D_F^*$ & Coefficient of molecular diffusion of pheromone & \SI{0.1}{cm^2/seg} & \cite{wyatt2003pheromones}\\\hline
	$\om^*$ & Mean angular speed & \SI{0.2}{rad/seg} & \cite{vela2015individual}  \\\hline
\end{tabular}}\label{Table_dimVars}\caption{Values of dimensional quantities used as baseline. The values of $\omega^*$ were computed from the values of the turning angles of ants in the laboratory measured every \SI{0.04}{seg} as reported in Table 1 of \cite{vela2015individual}. See also \cite{edelstein1995trail} for a useful compilation of parameter values of different species.} 
\end{table}


Note that Table \ref{Table_dimVars} contains no mention of a typical scale for the concentration of pheromone: that would require the knowledge of typical values of the rates $ f_A $ and $ f_N$. We circumvent this problem by tying the pheromone scale to that of the habitat. In particular, following \cite{wyatt2003pheromones} we assume there exists a concentration threshold $F_0$ (\si{mol/cm^3}) below which the ants cannot sense the pheromone. We could use an estimate of $ F_0 $ for a particular species of interest as the typical pheromone concentration scale, but instead we posit that if $ r^* $ is the maximum distance from the nest to which the ants can effectively forage, then 
\begin{equation}
F_N(\xb) = F_0 \;\; \text{for} \;\; \|\xb\|=r^*
\end{equation}
must hold. We then define the baseline pheromone concentration $F^*$ (\si{mol/cm^3}) as the average value of $F_N$ over the habitat:
\begin{equation}\label{Def_Fstar}
F^* = \frac{1}{\pi (r^*)^2} \int_{\|\xb\|<r^*} \!\! F_N(\xb) \ud \xb = \frac{f_N}{\pi  D_F \,  r^*} = 2 F_0.
\end{equation}
If both $ r^* $ and $ F_0 $ are known, then $ f_N $ can be obtained through \eqref{Def_Fstar}.

Setting the baseline time scale as $t^* = r^*/v^*$ yields a unitary adimensional speed. Further simplifications using the typical scales defined above result in the following non-dimensional version of the active walker model:
\begin{align}
	\ud {\xb}_j &=  (\cos \vp_j(t), \sin \vp_j(t)) \ud t \label{Eq_xdot_ND}\\
	\ud \vp_j &= \nu \, \omega(\xb_j,\vp_j) \ud t + \sqrt{2 D_\vp} \ud B_{\vp_j}, \label{Eq_dphiND}
\end{align}
for $ j=1,\dots,N $, coupled to the pheromone fields
\begin{align}
	F_N(\xb) &= \frac{1}{2 \| \xb \|}, \label{Eq_FN_ND}\\
	F_A(\xb,t) &= \sum_{i=1}^N \sum_{m=1}^{M_i(t)} \int_{T_{i,m}^F}^{T_{i,m}^N} f_A e^{-\kappa(s-T^F_{i,m})} G(\xb_i(s),\xb,t-s) \ud s \label{Eq_FA_ND}.
\end{align}
All variables and parameters in  (\ref{Eq_xdot_ND}-\ref{Eq_FA_ND}) are now unitless, but in order to keep notation light, we have used the same symbols as in the dimensional formulation (except for the new parameter $ \nu $, which is defined below). These adimensional variables and parameters can be written back in terms of their dimensional counterparts (now denoted with a bar) by:
\begin{equation}\label{Eq_NDvars}
	\begin{array}{cccc}
	\xb = \bar{\xb}/r^*, & t = \bar{t}/t^* , & \nu = \omega^* t^*,  \\[2ex]
	F_A = \bar{F}_A/F^*,  &   F_N = \bar{F}_N/F^*, & D_F = t^* D_F^*/r^{*2}, \\[2ex]
	  D_\vp = t^* \bar{D}_\vp,  &  f_A = \bar{f}_A/\bar{f}_N, & \kappa = t^* \bar{\kappa}.
	\end{array}
\end{equation}
We will use only non-dimensional quantities for the remaining of the paper with the hope that the change in notation does not create any confusion. In order to complete the model we must prescribe the function $ \omega $ and   appropriate ranges of values for $ \kappa, f_A $ and $D_\vp$. We do this in the next two sections.

\section{Tropotaxis, angular velocity and dispersion}\label{Sec_Tropotaxis}

We now attempt to model the phenomenon of tropotaxis via the angular velocity $ \omega $ in \eqref{Eq_dphiND}. The basic assumption is that an ant is able to sense only the gradient of pheromone concentration in order to make changes in orientation that will guide her in the direction of the highest pheromone gradient. This change of orientation is denoted by $ \ud \vp $ and modeled by the drift term $ \omega \ud t $ of the stochastic differential equation \eqref{Eq_dphiND}. 

Suppose that at time $t$, an ant occupies location $\xb$ while sensing a general pheromone signal $ F $. We propose the following decomposition of the angular velocity: 
\begin{equation}\label{Eq_omega}
\om(\xb,\vp) = \Omega[F](\xb,t) \,\, \tau(\vp,\grad F(\xb,t)).
\end{equation}
The term $ \Omega $ can be thought as the \textit{magnitude of the turn}: a functional of the local pheromone field $ F $ in terms of its gradient at the ant's location. We call  $ \tau $ the \textit{turn function} and it gives the direction of the orientation change as a function of the angle spanned by the gradient of the pheromone concentration and the ant's orientation, 
\[ \tau(\vp,\grad F(\xb,t)) = \tau(\measuredangle(\vp,\grad F(\xb,t))). \]

Which pheromone field $ F $ is evaluated in \eqref{Eq_omega} depends on the ants activity at time $t$. For foraging ants trying to locate food sources that may or may not have been located already, $ F= F_A $. Once an ant finds food and becomes a recruiter, we use either $ F = F_N $, or a combination of $ F_A $ and $ F_N $ as explained in Section \ref{Sec_trails}.

\subsection{The turn function $\tau$}

Let $\theta = \measuredangle (\vp,\grad F(\xb,t))$ stand for the angle of the gradient of pheromone concentration measured with respect to the ant's orientation. We use the convention in which angles are positive in the counterclockwise direction. Tropotaxis requires $\tau$ to be monotone, yield zero whenever the ant and the gradient of pheromone are aligned ($\theta=0$), and to give the maximum turn whenever the gradient points in precisely the opposite direction to the ant ($\theta = \pm \pi$). 

\begin{figure}\label{Fig_turnFunction}
\centering
\includegraphics[scale=0.9]{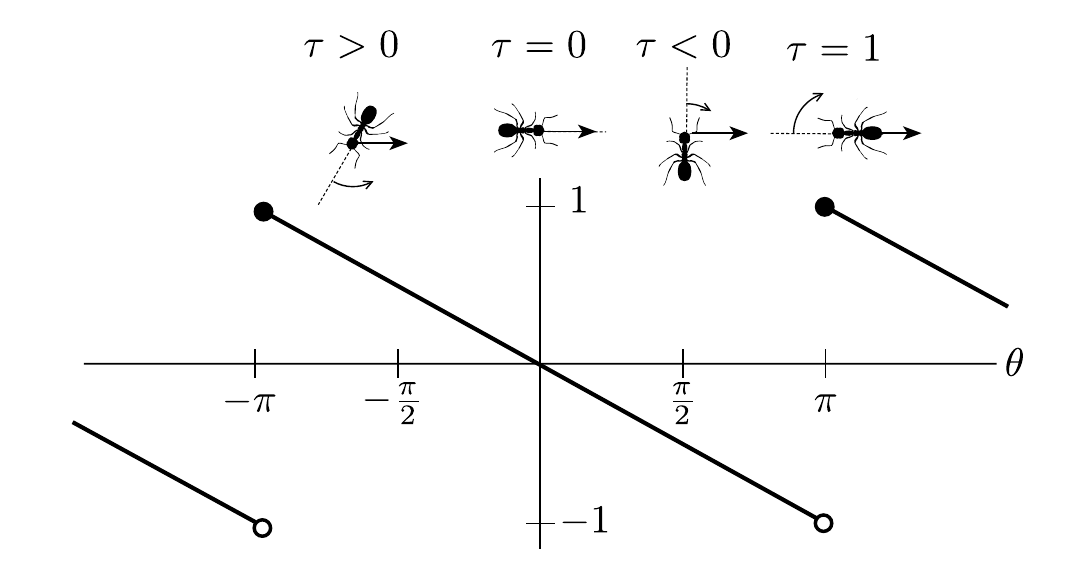}
\caption{Schematic definition of the turn function $\tau$: $\theta $ denotes the angle formed by the ants axis and $\grad F$, which is always pointing left in the diagram. The curved arrow indicates the direction of the corresponding turn $\ud \vp$.}
\end{figure}

Figure \ref{Fig_turnFunction} defines the simplest such function. The function is $2\pi$-periodic and can be taken right or left continuous depending on whether the ant turns to the right or the left in the presence of a gradient pointing in the direction opposite to hers. A useful representation for $\tau$ is
\begin{equation}\label{Eq_tau}
\tau(\theta) = -\sign(\sin(\theta)) \arccos(\cos(\theta)).
\end{equation}

\subsection{The magnitude of turns}\label{Sec_Omega}
Here we make considerations about how ants may take the sensory data of $F$ and $\grad F$ at a point, and convert it dynamically into modifications of their orientation via the functional $\Omega$ in \eqref{Eq_dphiND}. In \cite{perna2012individual} the authors propose that the ant's response to the pheromone field follows Weber's law: the turning angle is proportional to the difference between the concentration of the pheromone on both sides of their body, divided by the mean concentration. Mathematically one may write
\begin{equation}\label{Eq_WeberLaw}
\Omega_W[F] \propto \frac{\grad F}{F} = \grad(\log(F)).
\end{equation}
The appearance of the logarithm in \eqref{Eq_WeberLaw} is telling as it means that ants may be exposed to pheromone fields with concentrations varying over several orders of magnitude. That this is the case can be seen even from the nest pheromone field. Recall that the field $F_N$ in equation \eqref{Eq_FN_ND} was obtained by solving the steady three-dimensional linear molecular diffusion equation with a point source located at the origin. The resulting field is singular at the origin and radially symmetric, so we can write it $F_N= F_N(r)$ where $r$ is the distance to the nest. At the edge of the habitat $F_N(1) = 1/2$, and for $r\sim 10^{-3}$, comparable to an ants size, one has $F_N \sim 10^3$. We thus have a pheromone field that varies at least over three orders of magnitude within the habitat. We will see in Section \ref{Sec_recruitment} that this range is necessarily even wider in the case of the pheromone produced by ants. 

\textbf{Formula (26) is, similarly, shoehorned in to fit the expected behavior.}

Our proposal for the feedback mechanism encoded in $ \omega $ assumes pure tropotaxis, namely that ants can only sense the gradient of pheromone concentration. We posit
\begin{equation}\label{Def_OmegaT}
\Omega[F] = \log(1+|\grad{F}|),
\end{equation}
where the added unity prevents $ \Omega $ to become singular at locations of very low pheromone gradient (see equation (2.5) in \cite{amorim2018ant} for the corresponding correction in the case of the generalized Weber's law). 

\begin{figure}\centering
\includegraphics[width=\textwidth]{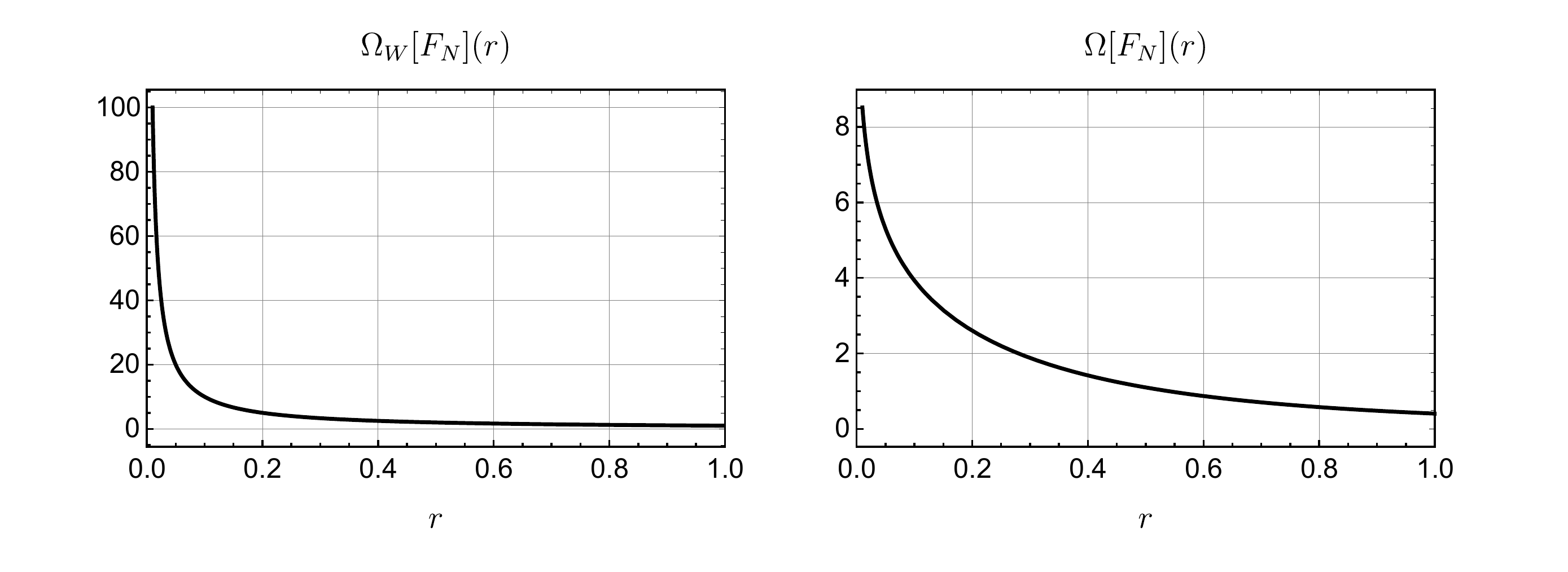}
\caption{Comparison of models for the magnitude of the turn in equations \eqref{Eq_WeberLaw} and \eqref{Def_OmegaT}.}\label{Fig_Omegas}
\end{figure}

Figure \ref{Fig_Omegas} shows the comparison between our proposal and Weber's law. Note that in both models $\Omega[F_N]$ becomes singular at the nest. This should pose no problem for modeling since in the close vicinity of the nest's entrance an ant may use other cues to orient herself and tropotaxis is likely irrelevant there. We thus introduce a small radius $\epsilon_N$ around the nest within which we assume our model does not apply. An estimate of $\epsilon_N$ will come from the analysis of successful recruitment conditions. 

\subsection{Exploration, return and angular dispersion}\label{Sec_Dispersion}
Consider the first ant that leaves the nest in search of food. Her foraging cycle can be divided into three stages:
\begin{enumerate}
	\item She leaves the nest and the ant pheromone field $ F_A $ is zero everywhere and so is the tropotaxis term affecting her orientation changes, $ \omega = 0$. According to our model, her success as a forager thus depends solely on the random variations in her orientation, namely $ \ud \vp =  \sqrt{2 D_\vp} \ud B_\vp $.
	\item When (and if) she finds a food source, her nest-bound journey will be guided by $ \om = \Omega[F_N]  \tau(\vp,\grad F_N) $ as angular velocity in the active walker equation, with $ \Omega $ and $ F_N $ given by \eqref{Def_OmegaT} and \eqref{Eq_FN_ND} respectively.
	\item After her return to the nest, she becomes a forager once again but will encounter the field $ F_A $ resulting from her (or any other successful scout's) pheromone markings, that is, $ \om = \Omega[F_A]  \tau(\vp,\grad F_A) $. 
\end{enumerate}

The angular dispersion coefficient $ D_\vp $ plays an important role in every phase of the cycle we just described. During the initial exploration phase, in which $ \omega =0 $, it can be shown that for constant $ D_\vp $, the mean dispersion rate is given by
\begin{equation}\label{Eq_dispRate}
\deriv{}{t} \EXP \|\xb^2(t)\| =  \frac{6 - 2 e^{-4 D_\vp t} - 4 e^{- D_\vp t} }{3 D_\vp}.
\end{equation}
Notice that \eqref{Eq_dispRate} exhibits ballistic motion (i.e. small dispersion) as $ t \downarrow 0 $ but quickly transitions to Fickian diffusion as $ t $ grows 
(see also \cite{romanczuk2012active}). We pose that such a switch in the dispersion rate behavior should not depend exclusively on time, but should instead also take into account the ant's activities and external stimuli, namely the pheromone fields. To be specific we propose the following: (i) while in close proximity to the nest, where $ F_N $ is large, the ant's dynamics should be ballistic. (ii) As an ant explores farther reaches away from the nest, the noise component of their motion should increase in order for her to improve the chances of finding food before leaving the region where $ F_N(\xb) \geq F_0$. 

\begin{figure}
	\centering
	\includegraphics[width=\textwidth]{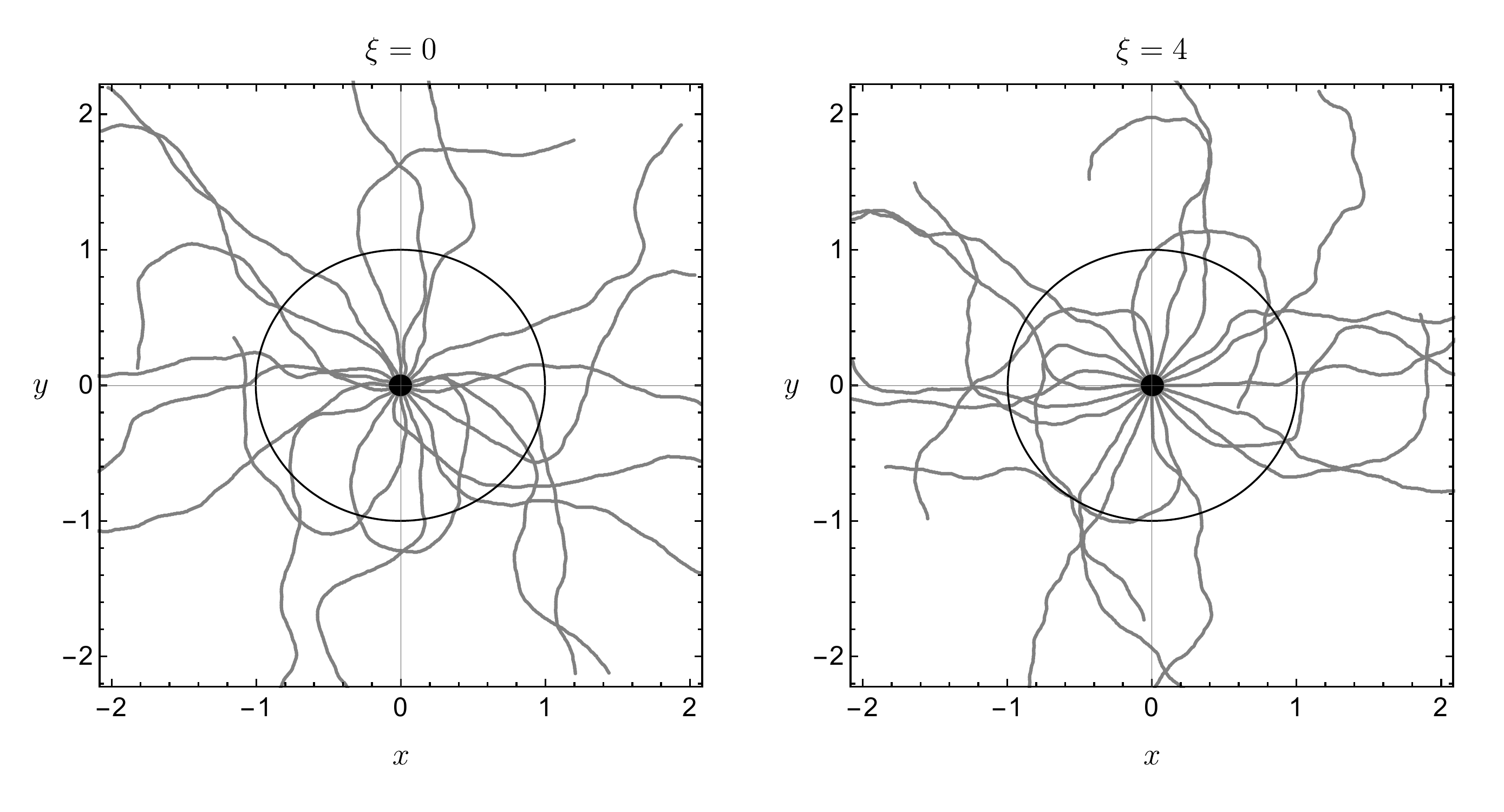}
	\caption{
	Simulations of twenty foraging ants initially located at the nest and facing outwards with equally spaced orientations. The angular dispersion is taken as \eqref{Eq_dispRate} with  $ \xi = 0 $, $ \xi = 4 $ in the left and right panels respectively, and $ D_\vp^{\max} = 10^{-1}t^*  $ in both cases. The simulations were run until $ t = 3 $. The black circle marks the habitat $ \|\xb \|\leq 1 $. No pheromone feedback is considered here. \label{Fig_explorationXi}} 
\end{figure}

We then propose that foraging ants experience a pheromone-dependent angular dispersion coefficient
\begin{equation}\label{Eq_Dphixi}
D_\vp(F_N) = \frac{D_{\vp}^{\max}}{1+ F_N^\xi}
\end{equation}
for some $ \xi > 1 $ and a maximum value $D_{\vp}^{\max} $ that, according to  \eqref{Eq_NDvars} and the laboratory results reported in \cite{vela2015individual}, is of order  $ t^* \times 10^{-2}$. Note that under this model, $ D_\vp(F_N(r)) $ is a sigmoid function of $ r $, and thus consistent with \eqref{Eq_dispRate}. Figure \ref{Fig_explorationXi} compares paths of foraging ants for the case of $ \xi = 0 $ (Fickian diffusion) and $ \xi =4 $. Note that in the latter case, foraging paths are essentially radial in the vicinity of the nest, and less likely to exit the habitat in the long term than in the case of simple Fickian diffusion.

\begin{figure}\centering
\includegraphics[width=0.5 \textwidth]{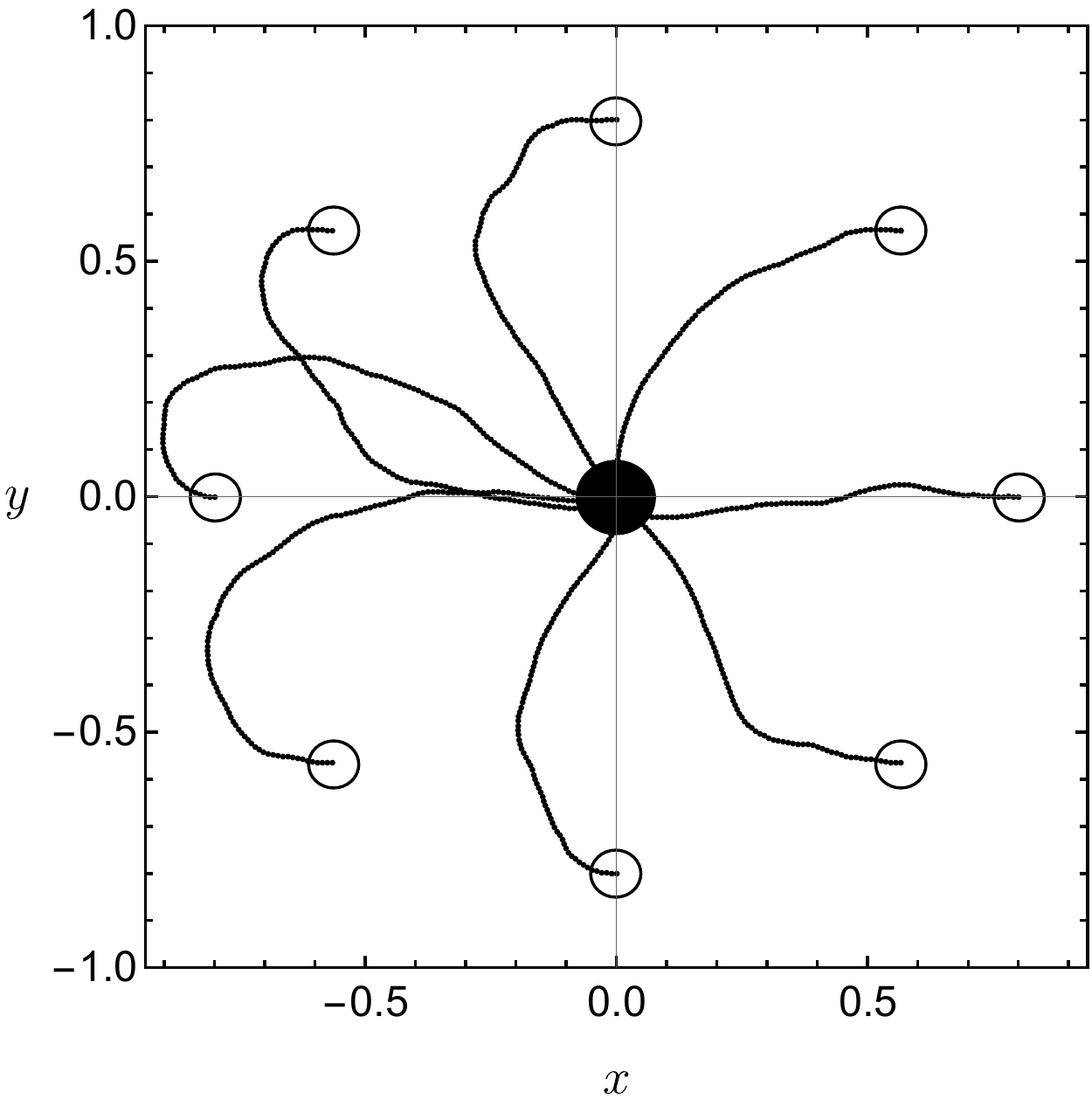}
\caption{Paths of eight ants as they return to the nest from different locations in the habitat. The initial position and orientation of each path are $\|\xb_j(0)\| = 0.8$ and $\vp_j(0) = \pi$ for all $ j $. See Remark \ref{Rem_return}}.\label{Fig_return2Nest}
\end{figure}

We assume that recruiting ants return to the nest aided only by the field $ F_N $ in \eqref{Eq_FN_ND}. Note that under our assumptions, the gradient $ \grad F_N(\xb)  = -\frac{1}{2} \|\xb\|^{-3/2} \xb$ gives, at each point, full information about the relative distance and direction of the nest. As long as ants are within the region $ \|\xb\| \leq 1 $, their return to the nest is easily accomplished by tropotaxis. Figure \ref{Fig_return2Nest} shows typical trajectories of the nest-bound journey of recruiting ants. 

\begin{remark}\label{Rem_return}
	Figure \ref{Fig_return2Nest} shows the importance of the initial orientation in the efficiency with which a recruiting ant can return to the nest according to equation \eqref{Eq_dphiND}. Any ant facing in the opposite direction of the nest, will have to undergo a long turn in order for the small changes $ \ud \vp $ to accumulate until it corrects its orientation. This is an artifact of the model, and in order to correct it, once an ant reaches a food source, we assume she can `turn around' and face the nest before starting her nest-bound journey. 
\end{remark}	

\section{Conditions for successful recruitment.}\label{Sec_recruitment}
We now turn our attention on the feedback mechanism between recruiting and foraging ants as mediated by the field $ F_A $. We thus focus on the active walker model (\ref{Eq_xdot_ND}-\ref{Eq_FA_ND}) for a foraging ant, namely $ \om = \Omega[F_A]  \tau(\vp,\grad F_A) $ where the field $ F_A $ is being continuously modified by a recruiting ant on her way to the nest. The goal of this section is to mathematically formulate  the successful recruitment conditions introduced in Section \ref{Sec_Model} and use them to arrive at acceptable values for the parameters $ f_A $ and $ \kappa $ in \eqref{Eq_FA_ND}. Our analysis here refines that of \cite{bossert1963analysis} where the authors consider only the case of constant emission rate $ g $.

Consider the $ j $-th ant in foraging mode, and suppose that at time $ T^F_{j,1} $ it finds a food source located at $ \xb_F $ with $ \|\xb_F\| <1 $. In order for her to be a successful recruiter, the parameters $ f_A$  and $ \kappa $  must be such that after the time $  T^N_{j,1} $, which is when the ant has returned to the nest, any other ant located at the nest can follow the direction of maximum gradient of the resulting field $ F_A $ and arrive at $ \xb_F $. Namely, we require the existence of a connected path $ \bs{p}:[T^N_{j,1},T^F] \to \R{^2} $ with $ \bs{p}(T^N_{j,1}) \approx (0,0) $ and $ \bs{p}(T^F) \approx \xb_F $ such that
\begin{equation}\label{Eq_SucRecruit}
	F_A(\bs{p}(t),t) > F_0, \quad \grad F_A(\bs{p}(t),t) \cdot \dot{\bs{p}}(t) > 0, \quad t \in [T^N_{j,1},T^F].
\end{equation}
Under these conditions, an ant would be able to walk along path $\bs{p}$ following an increasing concentration field and reach a vicinity of $\xb_F$ at time $T_F$. The fact that this path cannot exactly connect the nest at $(0,0)$ with $\xb_F$ will become apparent from the subsequent analysis.

In order to simplify the analysis, let us formulate a one-dimensional version of problem \eqref{Eq_SucRecruit}. Consider an ant that at time $t=0$ has just reached a food source at a distance $r_F\leq 1$ from the nest and is now in recruitment mode. We will now disregard all spatial variations of her path and assume she moves in a straight line with velocity $v$ back to the nest reaching her destination at time $t = r_F/v$. Along this path the ant would leave pheromone markings of time-varying strength $f_A e^{-\kappa t}$. It follows from \eqref{Eq_FA_ND} that the resulting pheromone field could be approximated for points along the path by
\begin{equation}\label{Eq_FA1D}
	F_A(r,t) = \int_0^{r_F/v} f_A e^{-\kappa s} G(r_F-sv,r,t-s) \ud s, \quad t\geq r_F/v
\end{equation}
where the distance to the nest $r$ is being used here as a proxy for a location along the straight line connecting the imaginary food source and the nest. The Green's function in \eqref{Def_G} is written in \eqref{Eq_FA1D} as
\begin{equation}\label{Def_G1D}
	G(r_F -sv,r,t-s) = \frac{1}{4(\pi D_F (t-s))^{3/2}}  \exp \left[- \frac{(r - 						r_F+sv)^2}{4 D_F (t-s)} \right].
\end{equation}
Our goal is to find conditions on $f_A$ and $\kappa$ such that (i) $F_A(r,t)$ is an increasing function of $r$, and (ii) $F(r,t) > F_0$ in an appropriately large portion of the interval $(0,r_F)$, for times $t\geq r_F/v$ and any $ r_F \leq 1 $.

There is no analytical expression that we know of for the integral in \eqref{Eq_FA1D}, however, key aspects of its behavior can be easily deduced. First note that $\pderiv{F_A}{r}(r_F,t) < 0$ for all $t > 0$. This is a consequence of diffusion: at the point where the recruiter starts leaving markings, there is a discontinuity in concentration that is instantly smoothed out by diffusion, thus creating a local maximum $r_1(\kappa,t) < r_F$. Close to the nest we have $\lim_{r\to 0^+} F_A(r,r_F/v) = + \infty$. Then, for a small period of time after $t = r_F/v$, diffusion forces $\pderiv{F_A}{r}(0,t)>0$ creating a local maximum and a local minimum close to the origin. See left panel of Figure \ref{Fig_FAFN}. Let $ r_2(\kappa,t) \geq 0$ denote the smallest local minimum. 

It is therefore clear that $F_A(r,t)$ is not a monotonically increasing function of $r$ on $(0,r_F)$ for all $t\geq r_F/v$. In practicality this means that no ant will be able to follow a trail from the nest until the food source by the sole mechanism of tropotaxis: they will get stuck at the local extremes of $F_A$. We circumvent this problem by assuming that there is a small radius $\epsilon_F>0$ such that if the ant is within a distance smaller than $\epsilon_F$ of the food, she will be able to locate it by some other means. We made a similar argument for introducing $\epsilon_N$ in Section \ref{Sec_Omega}. 

\begin{figure}[h!]\centering
\includegraphics[width=0.7\textwidth]{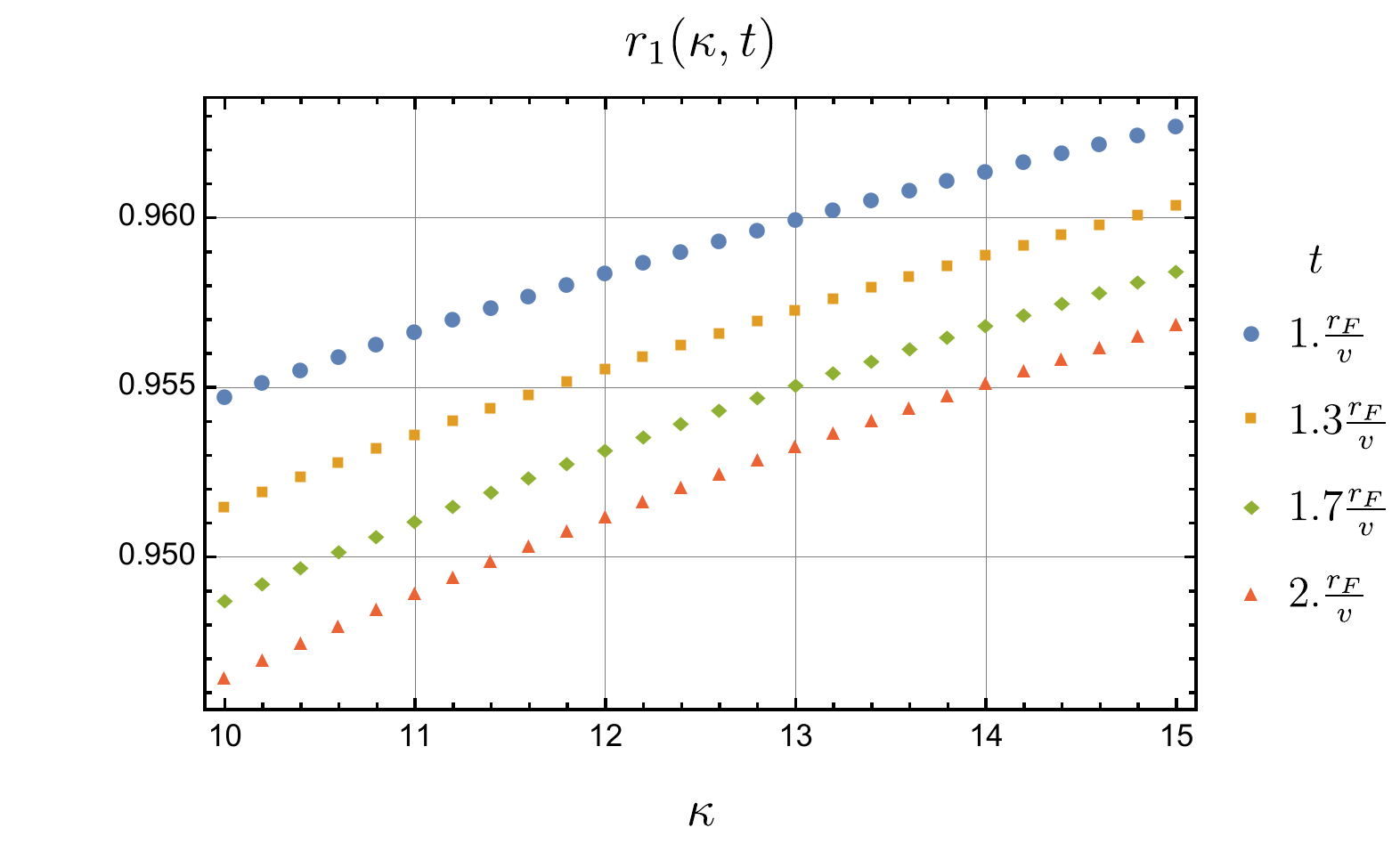}
\caption{Largest local maximum $r_1(\kappa,t)$ of $F_A(r,t)$ in equation \eqref{Eq_FA1D} as a function of $\kappa$ for selected times $t \geq r_F/v$. We used $r_F=1$, constant velocity $v=1$ and $D_F$ as in \eqref{Eq_NDvars}. If $\epsilon_F = 0.05$ and $\epsilon_N = 0.08$, condition \eqref{Eq_SRcondition} holds for $\kappa = 12$.}\label{Fig_r12}
\end{figure}

\begin{figure}[h!]\centering
\includegraphics[width=\textwidth]{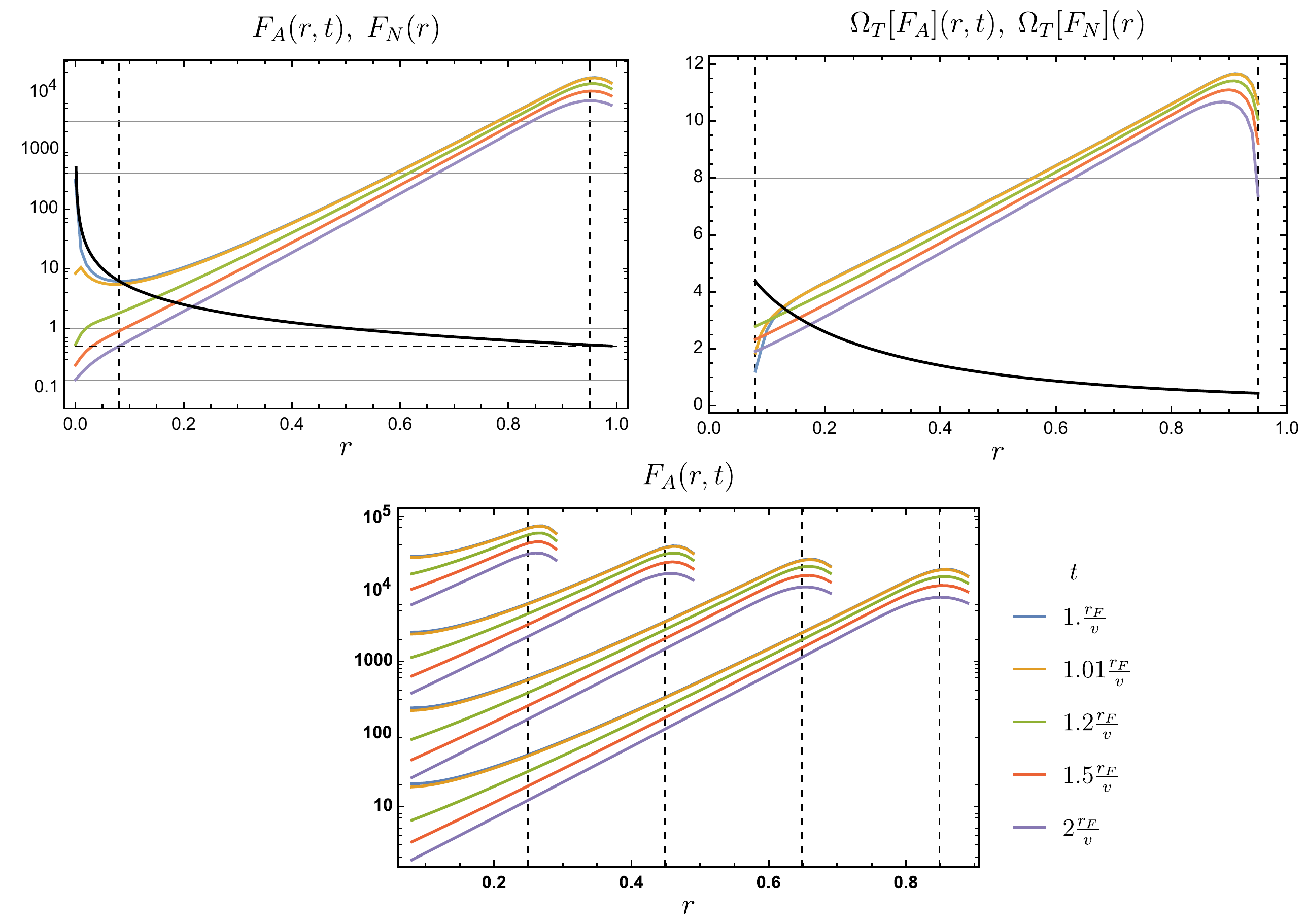}
\caption{Top figures: Log plots of $F_A(r,t)$ (left) and plots of $\Omega_T[F_A](r,t)$ (right) as a function of $r$ for selected values of $ t \in [r_F/v,2r_F/v]$ and $r_F=1$.  Vertical dashed lines mark $\epsilon_N$ and $r_F - \epsilon_F$, the threshold $F_0$ is marked by the horizontal dashed line. The values of $v,D_F,r_F,\kappa,\epsilon_N,\epsilon_F$ were chosen as in Figure \ref{Fig_r12}. We take $f_A = 60844$ as per \eqref{Eq_SR2}. For comparison, $F_N$ and $\Omega_T[F_N]$ are plotted in black. Bottom figure: Log plot of $F_A(r,t), \epsilon_N \leq r \leq r_F$ for selected times and four different values of the food location $r_F$; vertical dashed lines mark the values of $r_F - \epsilon_F$. }\label{Fig_FAFN}
\end{figure}

The first condition for successful recruitment can then be written as: given appropriate values for the tolerances $\epsilon_N$ and $\epsilon_F$, find values of $\kappa>0$ such that the locations $r_2$ and $r_1$ of the local extremes, satisfy
\begin{equation}\label{Eq_SRcondition}
	0\leq r_2(\kappa,t) < \epsilon_N < r_F-\epsilon_F < r_1(\kappa,t) < r_F, \quad \tfrac{r_F}{v} \leq t \leq \tfrac{2r_F}{v}.
\end{equation}
Condition \eqref{Eq_SRcondition} must be evaluated numerically, and guarantees a path of increasing pheromone concentration between distances $\epsilon_N$ and $r_F-\epsilon_F$ that lasts at least until any ant that was recruited in the nest at time $r_F/v$, subsequently reaches the food source. We observe that $ r_2(\kappa,t) = 0$ soon after $ t = r_F/v $, so $ r_2(\kappa,t) < \epsilon_N $ holds trivially. However, the behavior of $ r_1(\kappa,t) $ is monotone in both variables yielding a threshold value for $ \kappa $. See Figure \ref{Fig_r12}. 

Condition (ii) for successful recruitment requires the pheromone concentration $F_A(r,t)$ to be larger than the threshold $F_0$ for all distances and times of interest. Namely, 
\begin{equation}\label{Eq_SRcondition2}
f_A \geq \frac{F_0}{\int_0^{r_F/v} e^{-\kappa s} G(r_F-sv,r,t-s) \ud s}, \; \epsilon_N \leq r \leq r_F -\epsilon_F, \; \frac{r_F}{v} \leq t \leq \frac{2r_F}{v}.
\end{equation}
Note that the integral in the denominator is, by construction, an increasing function of $r$ in the interval $(\epsilon_N,r_F -\epsilon_F)$ and a monotonically decreasing function of $t$. It then suffices to take 
\begin{equation}\label{Eq_SR2}
	f_A \geq \frac{F_0}{\int_0^{r_F/v} e^{-\kappa s} G(r_F-sv,\epsilon_N, \frac{2r_F}{v}-s) \ud s}
\end{equation}
to ensure that the pheromone trail will be detectable until time $t = \frac{2r_F}{v}$. Figure \ref{Fig_FAFN} shows the resulting $F_A$ and $\Omega_T[F_A]$ as a function of $ r $, evidencing that the conditions \eqref{Eq_SRcondition} and \eqref{Eq_SRcondition2} for successful recruitment hold. Moreover,  the lower panel shows that successful recruitment conditions for $ r_F=1 $ guarantee the same for different values $ r_F \in[\epsilon_N, r_F] $ of the food location.

\begin{figure}[h]\centering
	\includegraphics[width=\textwidth]{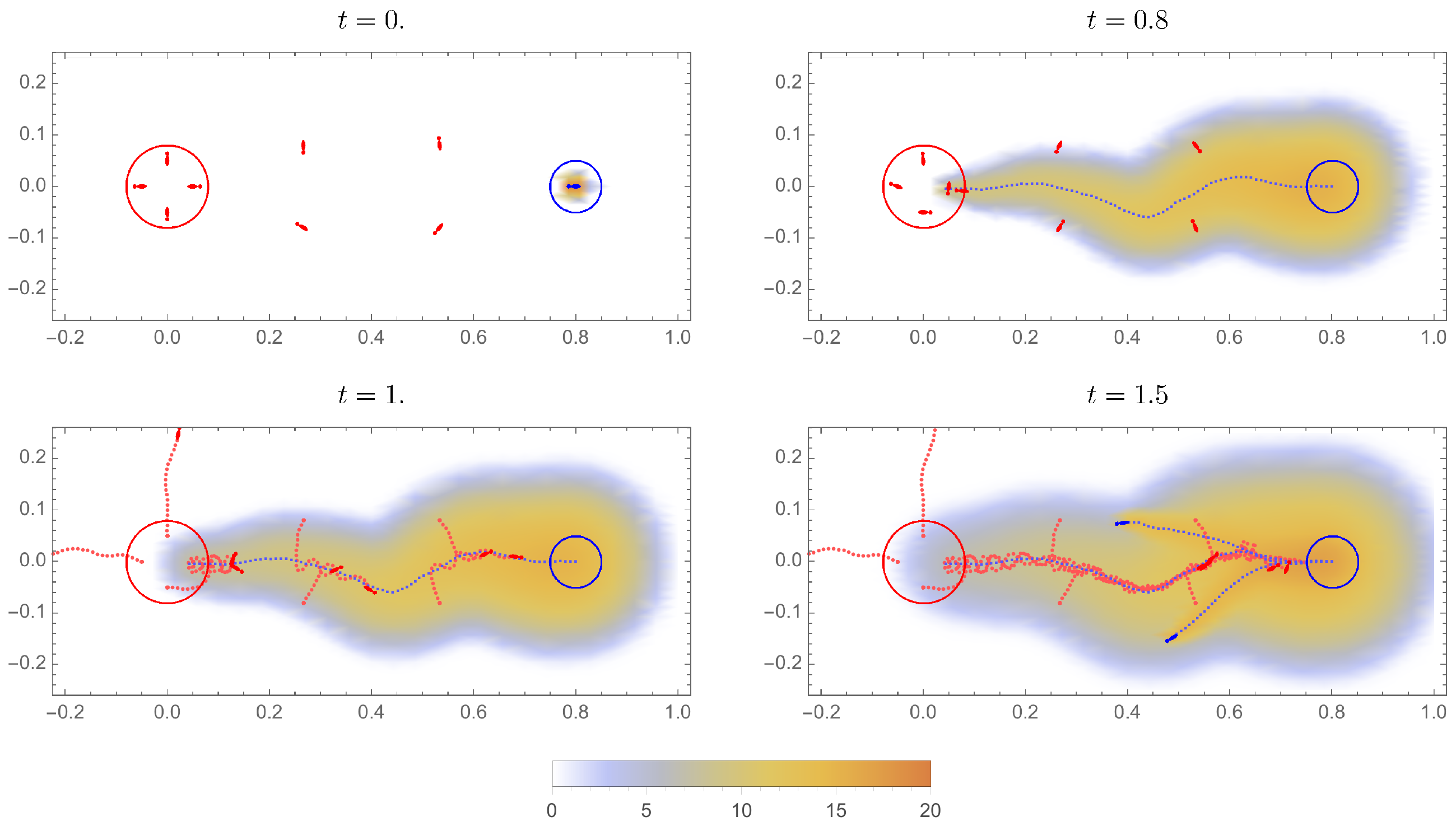}
	\caption{Four snapshots of a simulation with nine ants: one recruiter in blue originally at the food source ($r_F=0.8$), and eight red foragers in the nest and the vicinity of the food-nest trail. Each panel shows a density plot of $ \log(1+F_A(\xb,t)) $ (the $ \log $ is used here so $F_A$ can be visualized across all its range).  At $t=0$ the recruiter starts moving towards the nest attracted by $ F_N$. In this experiment, foragers are kept immobile until $ t=\frac{r_F}{v} = 0.8 $. At $ t=1 $ recruitment into the path can be seen for seven of the foraging ants, including the ant that originally reached the nest. At $t=1.5$, two ants are returning to the nest and a well-established trail along the original recruiter's path can be seen. The parameters of the model were taken as in Figure \ref{Fig_FAFN} and angular dispersion coefficient with $ \xi =4 $ in \eqref{Eq_Dphixi}}\label{Fig_Recruitment}
\end{figure}

The values of $ f_A $ and $ \kappa $ required for successful recruitment in the case of the parameter values in Table \ref{Table_dimVars} are $ f_A = 60844 $, $ \kappa = 12 $. Figure \ref{Fig_Recruitment} shows an example of successful recruitment using the appropriate parameters in the two-dimensional simulation. Notice the zig-zagging exhibited by foraging ants trying to follow the pheromone signal to the food source. This behavior has been documented and discussed in the context of tropotaxis in \cite{wilson1962chemical} and \cite{shorey1973behavioral}.

\section{Emergence of trail networks}\label{Sec_trails}
As it stands, our model predicts that once an ant finds a food source, she will return to the nest following the radially symmetric gradient of $F_N$. Hence, upon simulating several foraging cycles of many ants, a star-shaped collection of trails appears (see Figure \ref{Fig_return2Nest}). It has been widely observed however, that instead of a mere juxtaposition of individual routes, trails in large ant colonies often organize themselves into a network. \cite{deneubourg1990exploratory,traniello1991search,fourcassie1994dynamics,fourcassie2003dispersion}.

Although many environmental and behavioral factors are implicated in the specific topology of the resulting trail networks, it seems that the individual-based mechanism responsible for the appearance of this spatial patterns is a preference of recruiting ants to follow well-established trails to the nest instead of the shortest-path route \cite{holldobler1970recruitment,holldobler1990ants,holldobler1980foraging,detrain1991dynamics}. This preference can be included in our model by using the pheromone field $F_A$ as a secondary guidance signal for ants returning to the nest. Namely, we propose that for ants in recruitment mode, the turn function $ \tau $ in \eqref{Eq_omega} should compare the ants orientation with a convex combination of the gradients of $F_N$ and $F_A$:
\begin{equation}\label{Eq_omalpha}
\om(\xb_j,\vp_j) = \Omega[F_N] \,\,\, \tau\!\left(\vp_j,\alpha \frac{\grad F_N}{\| \grad F_N\| } + (1-\alpha) \frac{\grad F^{\neq j}_A}{\| \grad F^{\neq j}_A\| }\right),
\end{equation}
where $ 0\leq \alpha \leq 1 $ is a parameter, and $ F^{\neq j}_A $ is the pheromone field produced by all other ants except for the $ j $-th, namely the outer-most summation in \eqref{Eq_FA_ND} is taken here over all $ i \neq j $. The rationale for excluding the fraction of $ F_A $ produced by the $ j $-th ant is as follows.  She is actively marking the ground, and creating points of very high concentration gradients behind her. If these contributions were included for navigation purposes, the ant would tend to turn back around. This condition is similar to that obtained in \cite{amorim2018ant}: for an ant to stably follow a straight line in the direction of a pheromone gradient, her sensory area cannot include directions pointing to the rear of her movement.

\begin{figure}
	\includegraphics[width=\textwidth]{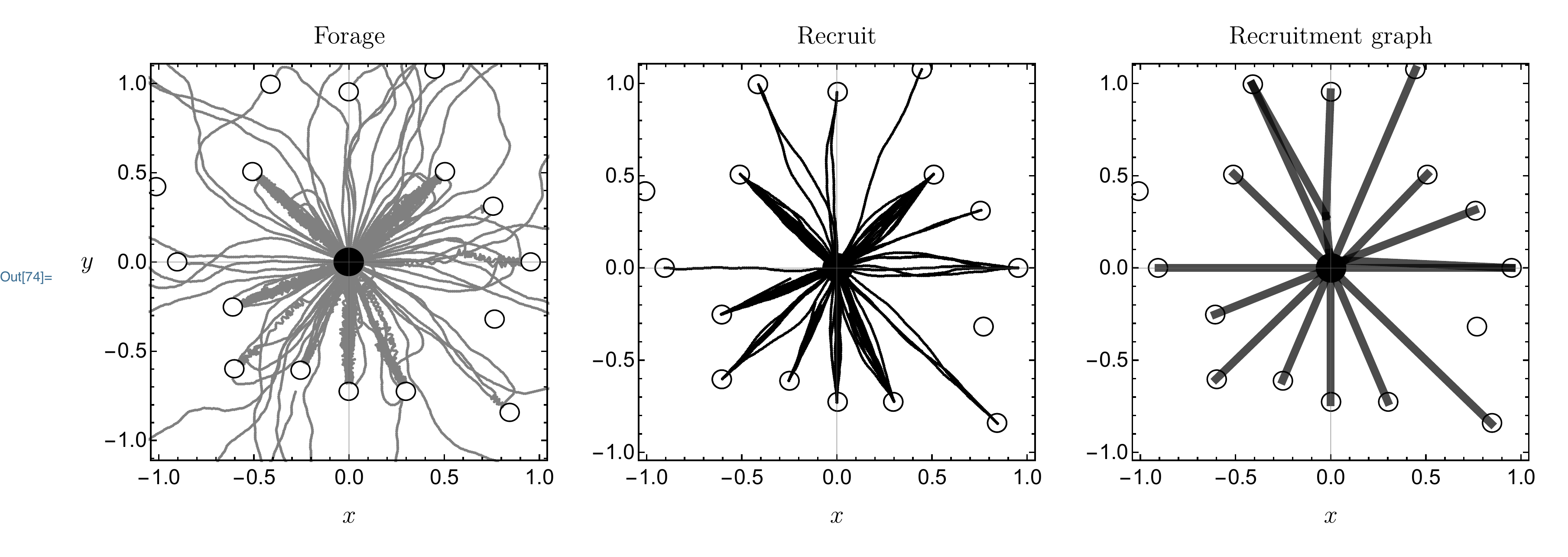}
	\caption{Simulation results of the foraging cycle of 100 ants for $ 0\leq t \leq 4 $ and sixteen food sources. The left panel shows the trajectories of ants during their foraging state while in the middle we show only trajectories of recruiting ants. The right panes shows a graph constructed by connecting the food sources, the nest, and approximate points of intersections between the trails in the middle panel. No trail preference is considered, which produces essential a star-graph of trails}\label{Fig_ForRecGraph}
\end{figure}

\begin{figure}
	\includegraphics[width=\textwidth]{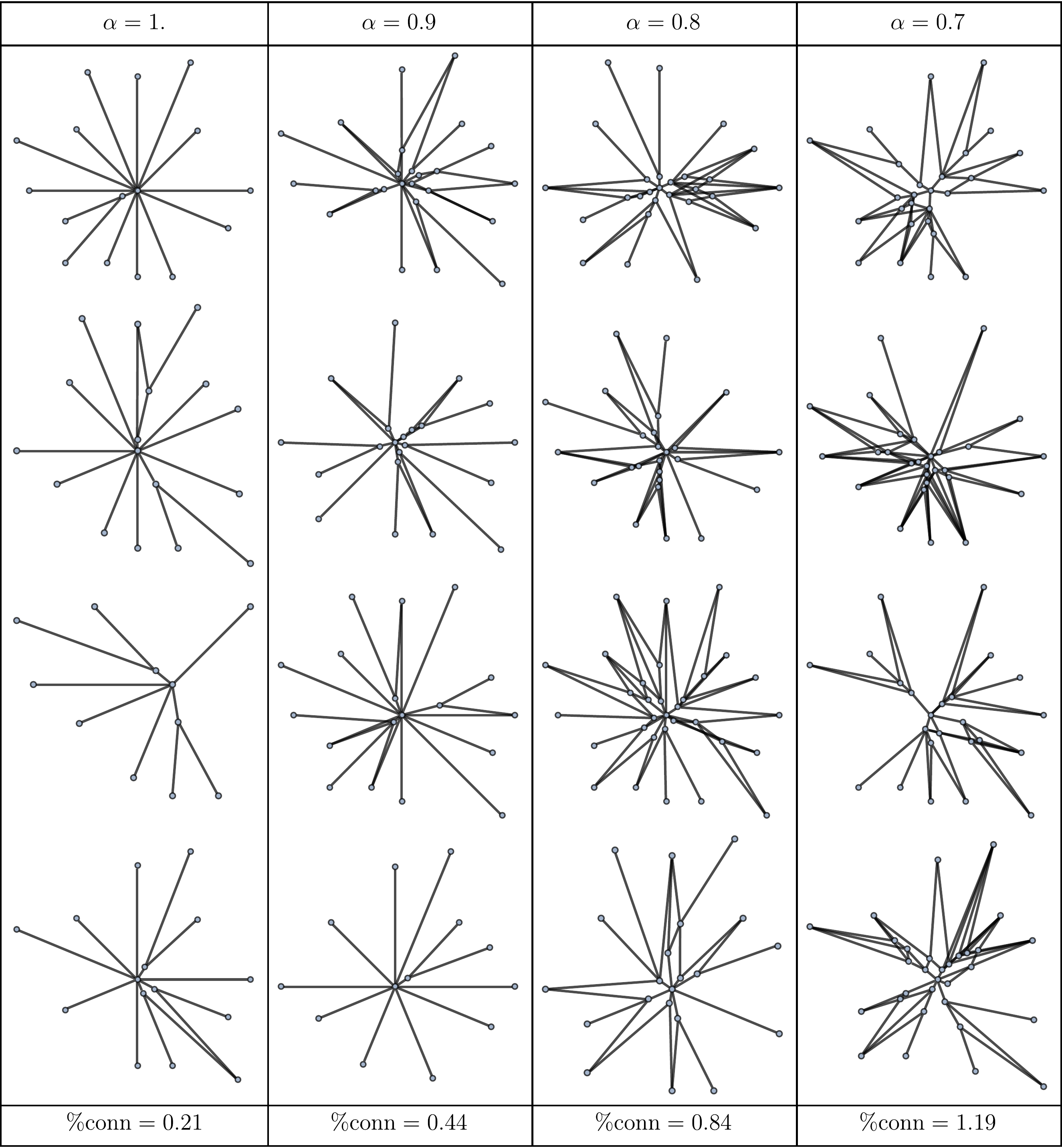}
	\caption{Reconstructed recruitment trail graphs from simulations of the foraging cycle for different values of the trail preference parameter $ \alpha $. Ten simulations (only five shown) were performed for each value of $ \xi $. All other parameters and initial conditions were kept constant. At the bottom of each case we report the mean, taken over the simulations of each case, of the fraction of vertices in the resulting graph that have degree larger than two.  }\label{Fig_allGraphs}
\end{figure}

We call $ \alpha $ the \textit{trail preference parameter}, and for $ \alpha =1$ we recover the model considered in Section \ref{Sec_recruitment}. As $ \alpha $ decreases, ants will tend to orient themselves where previous ants have laid trails. In order to explore the role played by trail preference, we used clustering algorithms to extract the emerging pattern of nest-bound trails arising from simulations of the ant foraging cycle. Figure \ref{Fig_ForRecGraph} shows the result of one simulation for $ \alpha =1 $ where it can be seen the emerging of a mostly radial trail network. Figure \ref{Fig_allGraphs} shows how as $ \alpha $ decreases from one, ants are more likely to follow existing trails on their nest-bound journey, and a more interconnected trail system emerges. The number of trail intersections, as measured as the fraction of nodes in the resulting graph with more than one incoming edge, decreases with increasing $ \alpha $. 

Another feature of colony-wide foraging that has been observed in several ant studies is the following: when multiple foods are available at the same time, the colony tends to exploit only a subset of them and travel along a few highly frequented trails \cite{pasteels1987self,deneubourg1989blind,detrain2008collective}. See also \cite{wyatt2003pheromones}. Namely, although several food sources can be discovered and pheromone markings laid guiding to those sources, the colony will tend to focus on a few food sources, traveling on the few corresponding trails, until those are exhausted. We will now illustrate that this behavior can be explained solely by tropotaxis.

\begin{figure}
	\includegraphics[width=\textwidth]{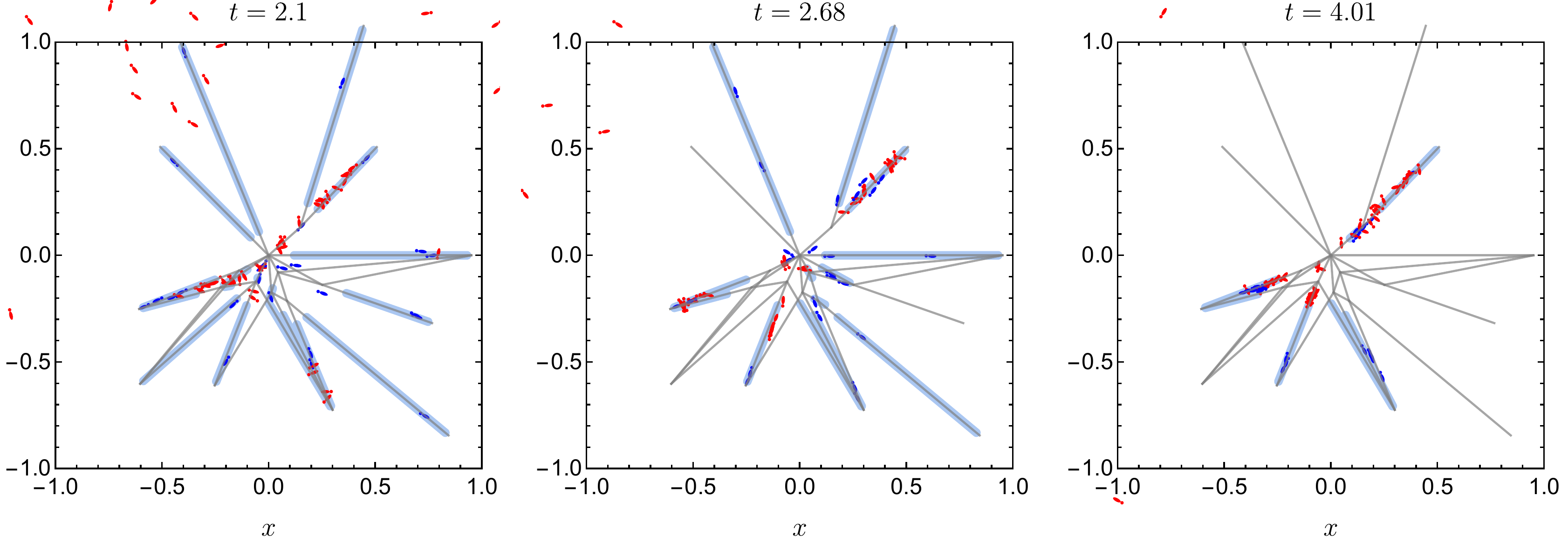}
	\caption{Evolution of trails used by ants for the case of $ \alpha =0.8 $ shown in the last row of Figure \ref{Fig_allGraphs}. Four different snapshots of the simulation are show. Light blue shows which trails are being transversed by recruiting ants (highlighted in bright blue, foragers in red). Very food source is maintained available during the whole simulation }\label{Fig_TrailHistory}
\end{figure}

In our simulations we do not consider the depletion (or quality) of food sources.  Hence all food sources are equally likely to be found, and the pheromone markings left by recruiters on their nest-bound trip have the same initial intensity $ f_A $. However, the number of ants traveling on the trails is far from static. Figure \ref{Fig_TrailHistory} shows the results of a simulation where, even though most food sources a quickly discovered by recruiters, by the end of the simulation all ants are focusing on only four of the available food sources and traveling along their corresponding trails. This behavior is observed regardless of the parameter $ \alpha $ and is a simple consequence of the trail-reinforcement feedback: the disproportional numbers of workers choosing and laying pheromones on the more traveled branch means that the ants stream to one source/path out-competes the stream to the other source/path \cite{goss1989self,detrain2008collective}.

\section{Summary and Discussion}

We have built an active walker model for the foraging/recruiting cycle of individual ants. The model explicitly tracks the spatio-temporal structure of the pheromone field produced by each ant, and how it feedbacks upon the dynamics of ants' orientations. The key assumption is that this feedback is mediated solely by tropotaxis in the form of the operator $ \Omega $ in \eqref{Def_OmegaT}

The model is not intended as a predictor of ant behavior. It may be used, however, to glimpse at qualitative properties of ant communication required for the emergence of observed colony-wide features of foraging. We show that under suitable assumptions on its parameters, our model can explain key aspects of the foraging/recruiting cycle of ants: searching for and finding food, returning food to the nest, recruiting ants to food sources and preference for movement along trails. Moreover, at the colony-wide scale, our model predicts the emergence of a trail network centered around the nest, with a degree of connectivity that is a function of the tropotaxis feedback. Finally, as shown in Figure \ref{Fig_TrailHistory}, tropotaxis seems to sufficient explanation for the focusing of the colony on few food sources/trails as several foraging cycles are completed.

The analysis leading to the conditions of successful recruitment in Section \ref{Sec_recruitment} provides a mathematical answer to the following question: if tropotaxis is the sole mode of communication, what do individual ants have to be able to do in order to successfully forage as a colony? 

We have found that the rigors of 3-dimensional diffusion impose serious constraints on the sensory and pheromone marking abilities of any ant colony depending on tropotaxis for recruitment. Figure \ref{Fig_FAFN} shows that ants need to accurately differentiate a  pheromone field with concentrations spanning five orders of magnitude. Moreover, the mechanism by which the gradient in pheromone concentration is converted into changes of orientation, must be tuned to operate over an input that varies over nine orders of magnitude. Similarly, the value of the exponent $ \kappa \sim 10$ in the marking rate \eqref{Def_gj} implies that over a typical return to the nest, ants must be able to make precise pheromone markings at rates encompassing six orders of magnitude. All these would be remarkable abilities and, as far as we know, have not been tested in the field. Note that these qualitative conclusions do not depend on the choice of base values in Table \ref{Table_dimVars}, they follow from a strict application of the dynamics of 3-dimensional diffusion and our choice of form for the tropotactic feedback.


\bibliographystyle{spmpsci}      
\bibliography{Hormigas}

\begin{thebibliography}{10}
\providecommand{\url}[1]{{#1}}
\providecommand{\urlprefix}{URL }
\expandafter\ifx\csname urlstyle\endcsname\relax
  \providecommand{\doi}[1]{DOI~\discretionary{}{}{}#1}\else
  \providecommand{\doi}{DOI~\discretionary{}{}{}\begingroup
  \urlstyle{rm}\Url}\fi

\bibitem{amorim2015modeling}
Amorim, P.: Modeling ant foraging: a chemotaxis approach with pheromones and
  trail formation.
\newblock Journal of theoretical biology \textbf{385}, 160--173 (2015)

\bibitem{amorim2018ant}
Amorim, P., Goudon, T., Peruani, F.: An ant navigation model based on weber's
  law  (2018)

\bibitem{beckers1992trails}
Beckers, R., Deneubourg, J.L., Goss, S.: Trails and u-turns in the selection of
  a path by the ant \textit{Lasius niger}.
\newblock Journal of theoretical biology \textbf{159}(4), 397--415 (1992)

\bibitem{boissard2013trail}
Boissard, E., Degond, P., Motsch, S.: Trail formation based on directed
  pheromone deposition.
\newblock Journal of mathematical biology \textbf{66}(6), 1267--1301 (2013)

\bibitem{bossert1963analysis}
Bossert, W.H., Wilson, E.O.: The analysis of olfactory communication among
  animals.
\newblock Journal of theoretical biology \textbf{5}(3), 443--469 (1963)

\bibitem{calenbuhr1992model}
Calenbuhr, V., Deneubourg, J.L.: A model for osmotropotactic orientation (i).
\newblock Journal of theoretical biology \textbf{158}(3), 359--393 (1992)

\bibitem{camazine2003self}
Camazine, S., Deneubourg, J.L., Franks, N.R., Sneyd, J., Bonabeau, E.,
  Theraula, G.: Self-organization in biological systems, vol.~7.
\newblock Princeton University Press (2003)

\bibitem{cheung2009mathematical}
Cheung, A.: Mathematical and neural network models of medium-range navigation
  during social insect foraging.
\newblock Food Exploitation By Social Insects: Ecological, Behavioral, and
  Theoretical Approaches p. 293 (2009)

\bibitem{deneubourg1990exploratory}
Deneubourg, J.L., Aron, S., Goss, S., Pasteels, J.M.: The self-organizing
  exploratory pattern of the argentine ant.
\newblock Journal of Insect Behavior \textbf{3}(2), 159--168 (1990).
\newblock \doi{10.1007/BF01417909}

\bibitem{deneubourg1989blind}
Deneubourg, J.L., Goss, S., Franks, N., Pasteels, J.: The blind leading the
  blind: modeling chemically mediated army ant raid patterns.
\newblock Journal of insect behavior \textbf{2}(5), 719--725 (1989)

\bibitem{detrain2008collective}
Detrain, C., Deneubourg, J.L.: Collective decision-making and foraging patterns
  in ants and honeybees.
\newblock Advances in insect physiology \textbf{35}, 123--173 (2008)

\bibitem{detrain1991dynamics}
Detrain, C., Deneubourg, J.L., Goss, S., Quinet, Y.: Dynamics of collective
  exploration in the ant pheidole pallidula.
\newblock Psyche: A Journal of Entomology \textbf{98}(1), 21--31 (1991)

\bibitem{ebeling2003active}
Ebeling, W., Schweitzer, F.: Self-organization, active brownian dynamics,and
  biological applications.
\newblock Nova Acta Leopoldina \textbf{88}, 169--188 (2003)

\bibitem{edelstein1995trail}
Edelstein-Keshet, L., Watmough, J., Ermentrout, G.B.: Trail following in ants:
  individual properties determine population behaviour.
\newblock Behavioral Ecology and Sociobiology \textbf{36}(2), 119--133 (1995)

\bibitem{fourcassie2003dispersion}
Fourcassi{\'e}, V., Bredard, C., Volpatti, K., Theraulaz, G.: Dispersion
  movements in ants: spatial structuring and density-dependent effects.
\newblock Behavioural processes \textbf{63}(1), 33--43 (2003)

\bibitem{fourcassie1994dynamics}
Fourcassie, V., Deneubourg, J.L.: The dynamics of collective exploration and
  trail-formation in \textit{Monomorium pharaonis}: experiments and model.
\newblock Physiological Entomology \textbf{19}(4), 291--300 (1994)

\bibitem{goss1989self}
Goss, S., Aron, S., Deneubourg, J.L., Pasteels, J.M.: Self-organized shortcuts
  in the argentine ant.
\newblock Naturwissenschaften \textbf{76}(12), 579--581 (1989)

\bibitem{hangartner1967spezifitat}
Hangartner, W.: Spezifit{\"a}t und inaktivierung des spurpheromons von lasius
  fuliginosus latr. und orientierung der arbeiterinnen im duftfeld.
\newblock Zeitschrift f{\"u}r vergleichende Physiologie \textbf{57}(2),
  103--136 (1967)

\bibitem{helbing1997active}
Helbing, D., Schweitzer, F., Keltsch, J., Moln{\'a}r, P.: Active walker model
  for the formation of human and animal trail systems.
\newblock Physical review E \textbf{56}(3), 2527 (1997)

\bibitem{holldobler1980foraging}
H{\"o}lldobler, B., M{\"o}glich, M.: The foraging system of \textit{Pheidole
  militicida} (hymenoptera, formicidae).
\newblock Insectes Sociaux \textbf{27}(3), 237--264 (1980)

\bibitem{holldobler1970recruitment}
H{\"o}lldobler, B., Wilson, E.: Recruitment trails in the harvester ant,
  pogonomyrmex badius.
\newblock Psyche \textbf{77}(4), 385--399 (1970)

\bibitem{holldobler1990ants}
H{\"o}lldobler, B., Wilson, E.O.: The ants.
\newblock Harvard University Press (1990)

\bibitem{jackson2004trail}
Jackson, D.E., Holcombe, M., Ratnieks, F.L.: Trail geometry gives polarity to
  ant foraging networks.
\newblock Nature \textbf{432}(7019), 907 (2004)

\bibitem{lam1995active}
Lam, L.: Active walker models for complex systems.
\newblock Chaos, Solitons \& Fractals \textbf{6}, 267--285 (1995)

\bibitem{leuthold1975orientation}
Leuthold, R.: Orientation mediated by pheromones in social insects.
\newblock Pheromones and Defensive Secretions in Social Insects pp. 197--211
  (1975)

\bibitem{pasteels1987self}
Pasteels, J.M., Deneubourg, J.L., Goss, S.: Self-organization mechanisms in ant
  societies. i. trail recruitment to newly discovered food sources.
\newblock In: From individual to collective behavior in social insects: les
  Treilles Workshop/edited by Jacques M. Pasteels, Jean-Louis Deneubourg.
  Basel: Birkhauser, 1987. (1987)

\bibitem{perna2012individual}
Perna, A., Granovskiy, B., Garnier, S., Nicolis, S.C., Lab{\'e}dan, M.,
  Theraulaz, G., Fourcassi{\'e}, V., Sumpter, D.J.: Individual rules for trail
  pattern formation in argentine ants (\textit{Linepithema humile}).
\newblock PLoS computational biology \textbf{8}(7), e1002,592 (2012)

\bibitem{polyanin2015handbook}
Polyanin, A.D., Nazaikinskii, V.E.: Handbook of linear partial differential
  equations for engineers and scientists.
\newblock Chapman and hall/crc (2015)

\bibitem{romanczuk2012active}
Romanczuk, P., B{\"a}r, M., Ebeling, W., Lindner, B., Schimansky-Geier, L.:
  Active brownian particles.
\newblock The European Physical Journal Special Topics \textbf{202}(1), 1--162
  (2012)

\bibitem{ryan2016model}
Ryan, S.D.: A model for collective dynamics in ant raids.
\newblock Journal of mathematical biology \textbf{72}(6), 1579--1606 (2016)

\bibitem{schienbein1993langevin}
Schienbein, M., Gruler, H.: Langevin equation, fokker-planck equation and cell
  migration.
\newblock Bulletin of Mathematical Biology \textbf{55}(3), 585--608 (1993)

\bibitem{schweitzer2007brownian}
Schweitzer, F.: Brownian agents and active particles: collective dynamics in
  the natural and social sciences.
\newblock Springer (2007)

\bibitem{schweitzer1997active}
Schweitzer, F., Lao, K., Family, F.: Active random walkers simulate trunk trail
  formation by ants.
\newblock BioSystems \textbf{41}(3), 153--166 (1997)

\bibitem{shorey1973behavioral}
Shorey, H.H.: Behavioral responses to insect pheromones.
\newblock Annual review of entomology \textbf{18}(1), 349--380 (1973)

\bibitem{sudd2013behavioural}
Sudd, J.H., Franks, N.R.: The behavioural ecology of ants.
\newblock Springer Science \& Business Media (2013)

\bibitem{taktikos2012collective}
Taktikos, J., Zaburdaev, V., Stark, H.: Collective dynamics of model
  microorganisms with chemotactic signaling.
\newblock Physical Review E \textbf{85}(5), 051,901 (2012)

\bibitem{traniello1991search}
Traniello, J., Fourcassi{\'e}, V., Graham, T.: Search behavior and foraging
  ecology of the ant \textit{Formica schaufussi}: colony-level and individual
  patterns.
\newblock Ethology Ecology \& Evolution \textbf{3}(1), 35--47 (1991)

\bibitem{vakeroudis2015windings}
Vakeroudis, S.: On the windings of complex-valued ornstein--uhlenbeck processes
  driven by a brownian motion and by a stable process.
\newblock Stochastics An International Journal of Probability and Stochastic
  Processes \textbf{87}(5), 766--793 (2015)

\bibitem{vela2015individual}
Vela-P{\'e}rez, M., Fontelos, M.A., Garnier, S.: From individual to collective
  dynamics in argentine ants (linepithema humile).
\newblock Mathematical biosciences \textbf{262}, 56--64 (2015)

\bibitem{wilson1962chemical}
Wilson, E.O.: Chemical communication among workers of the fire ant
  \textit{Solenopsis saevissima} (fr. smith) 1. the organization of
  mass-foraging.
\newblock Animal behaviour \textbf{10}(1-2), 134--147 (1962)

\bibitem{wyatt2003pheromones}
Wyatt, T.D.: Pheromones and animal behaviour: communication by smell and taste.
\newblock Cambridge university press (2003)

\end{thebibliography}

\end{document}